\documentclass[sigplan,screen]{acmart}

\usepackage{pgfplots}
\usepackage{tikz}
\usetikzlibrary{shapes,shapes.geometric,fit,positioning,calc,arrows}
\usepackage{amsmath}
\usepackage{color}
\usepackage[T1]{fontenc}
\usepackage{hyperref}
\usepackage[latin1]{inputenc}
\usepackage{balance}
\usepackage{subcaption}
\usepackage[capitalize]{cleveref}
\usepackage{listings}
\usepackage{graphicx}
\usepackage{mathpartir}
\usepackage{xspace}
\usepackage{array}
\usepackage{enumitem}
\usepackage{IEEEtrantools}
\usepackage{pifont}
\usepackage{booktabs}
\usepackage{xfrac}
\usepackage{makecell}
\usepackage{wrapfig}
\usepackage[vlined]{algorithm2e}
\usepackage{stmaryrd}
\usepackage{galois}
\usepackage[normalem]{ulem}
\usepackage{apxproof}
\usepackage{stfloats}
\usepackage{calc}
\usepackage{microtype}

\newtheoremrep{lemma}{Lemma}[section]
\newtheoremrep{theorem}{Theorem}[section]
\definecolor{darkred}{rgb}{0.75, 0, 0}
\definecolor{darkblue}{rgb}{0,    0, 0.75}
\definecolor{grey}{rgb}{0.65, 0.65, 0.65}
 \newcommand{\todo}[1]{\textcolor{darkred}{$\blacktriangleright$#1$\blacktriangleleft$}}
\newcommand{\ms}[1]{\todo{#1 --MS}}
\newcommand{\jedi}[1]{\textcolor{darkblue}{\begin{quotation}\textit{#1}\end{quotation}}}

\newcommand{\tightpara}[1]{\vspace{0.5em}\textbf{\emph{#1}}}

\newcommand{\ldenote}{\llbracket}
\newcommand{\rdenote}{\rrbracket}
\newcommand{\denote}[1]{\ldenote #1 \rdenote}

\newcommand{\cfedge}[3]{#1\!-\hspace{-0.6em} \mbox{-}\hspace{-0.18em} [#2]\hspace{-0.5em}\shortrightarrow\! #3}
\newcommand{\nameprod}{\!\cdot\!}
\newcommand{\daigeval}[6]{#1, #2 \vdash #3 \Rightarrow #4 ~; #5,#6}
\newcommand{\daigevaltemplate}{\daigeval{\mathcal D}{M}{n}{v}{\mathcal D'}{M'}}
\newcommand{\daigedit}[4]{#1\vdash #2 \Leftarrow #3 ~; #4}
\newcommand{\daigedittemplate}{\daigedit{\mathcal D}{n}{v_\varepsilon}{\mathcal D'}}

\newcommand{\aitemplate}{\langle \Sigma^\sharp,\varphi_0,\denote\cdot^\sharp,\sqsubseteq,\sqcup,\nabla\rangle}
\newcommand{\code}[1]{\texttt{\small #1}}
\newcommand{\fnname}{\underline f \nameprod (\underline {v_1} \cdots\underline {v_k})}

\makeatletter
\newcount\lst@savelstnumber
\newcommand{\lstnonum}{%
  \global\lst@savelstnumber\c@lstnumber%
  \gdef\thelstnumber{}%
}
\newcommand{\lststopn}{%
  \lstnonum%
}
\newcommand{\lstnumprefix}[1]{$\ell_{#1}$}
\newcommand{\lststartn}{%
  \global\c@lstnumber\lst@savelstnumber%
  \gdef\thelstnumber{\lstnumprefix{\arabic{lstnumber}}}%
}
\makeatother

\newcommand{\lstbasicstyle}{\ttfamily\fontseries{l}\selectfont}
\newcommand{\lstkeywordstyle}{\relsize{-0.5}\fontseries{b}\selectfont}
\newcommand{\lstcommentstyle}{\sl}
\newcommand{\lstnumberstyle}{\scriptsize\em}

\lstdefinestyle{default}{%
  backgroundcolor=\color{white},%
  basicstyle=\lstbasicstyle,%
  commentstyle=\lstcommentstyle,%
  keywordstyle=\lstkeywordstyle,%
  columns=fullflexible,%
  keepspaces=true,%
  mathescape%
}
\lstdefinestyle{number}{%
  numbers=left,%
  numberstyle=\lstnumberstyle,%
  xleftmargin=2em%
}

\lstdefinelanguage{jsimp}{
  keywords={function,if,return,while,null,var,assume,assert}
}

\setcopyright{rightsretained}
\acmPrice{}
\acmDOI{10.1145/3453483.3454044}
\acmYear{2021}
\copyrightyear{2021}
\acmSubmissionID{pldi21main-p111-p}
\acmISBN{978-1-4503-8391-2/21/06}
\acmConference[PLDI '21]{Proceedings of the 42nd ACM SIGPLAN International Conference on Programming Language Design and Implementation}{June 20--25, 2021}{Virtual, Canada}
\acmBooktitle{Proceedings of the 42nd ACM SIGPLAN International Conference on Programming Language Design and Implementation (PLDI '21), June 20--25, 2021, Virtual, Canada}

\usepackage{booktabs}   
\usepackage{subcaption} 


\makeatletter
\newcommand\footnoteref[1]{\protected@xdef\@thefnmark{\ref{#1}}\@footnotemark}
\makeatother

\begin{document}
\newsavebox{\SBoxAssume}\sbox{\SBoxAssume}{\footnotesize\lstinline{assume}}
\newsavebox{\SBoxNull}\sbox{\SBoxNull}{\footnotesize\lstinline{null}}

\title{Demanded Abstract Interpretation (Extended Version)}         


\author{Benno Stein}
\affiliation{
  \institution{University of Colorado Boulder}            
}
\email{benno.stein@colorado.edu}          

\author{Bor-Yuh Evan Chang}
\authornote{Bor-Yuh Evan Chang holds concurrent appointments at the University of Colorado Boulder and as an Amazon Scholar.  This paper describes work performed at CU Boulder and is not associated with Amazon.}
\orcid{0000-0002-1954-0774}             
\affiliation{
  \institution{University of Colorado Boulder}           
  \institution{Amazon}
}
\email{evan.chang@colorado.edu}         

\author{Manu Sridharan}
\orcid{0000-0001-7993-302X}             
\affiliation{
  \institution{University of California, Riverside}           
}
\email{manu@cs.ucr.edu}         



\newcommand{\demandedpninetyfive}{1.2}
\newcommand{\batchatdemandedpninetyfive}{\todo{X}}

\begin{abstract}
  We consider the problem of making expressive static analyzers \emph{interactive}.
  Formal static analysis is seeing increasingly widespread adoption as a tool for verification and bug-finding, but even with powerful cloud infrastructure it can take minutes or hours to get \emph{batch} analysis results after a code change.
  While existing techniques offer some demand-driven \emph{or} incremental aspects for certain classes of analysis, the fundamental challenge we tackle is doing {\em both} for arbitrary abstract interpreters.

  Our technique, {\em demanded abstract interpretation},
  lifts program syntax and analysis state to a dynamically evolving graph structure, in which
  program edits, client-issued queries, and evaluation of abstract semantics are all treated uniformly.
  The key difficulty addressed by our approach is the application of general incremental computation techniques to the complex, cyclic dependency structure induced by abstract interpretation of loops with widening operators.
  We prove that desirable abstract interpretation meta-prop\-er\-ties, including soundness and termination, are preserved in our approach, and that demanded analysis results are equal to those computed by a batch abstract interpretation.
  Experimental results suggest promise for a prototype demanded abstract interpretation framework:
  by combining incremental and demand-driven techniques, our framework consistently delivers analysis results at interactive speeds,
  answering 95\% of queries within \demandedpninetyfive~seconds.
\end{abstract}


\maketitle

\section{Introduction}\label{sec:intro}



Static analysis is seeing increasing real-world adoption for verification and bug finding, particularly as part of continuous integration (CI) and code review processes \cite{DBLP:conf/nfm/CalcagnoD11,DBLP:journals/cacm/SadowskiAEMJ18}.  However, a pain point with these deployments is that developers cannot get quick local analysis results for code they are editing; ideally, updated results would appear nearly instantly in their IDE.
%
In this paper, we present an {\em interactive} analysis engine designed to handle local queries and edits efficiently,
which can complement a {\em batch} engine that exhaustively analyzes a fixed program, for example by quickly verifying whether a local change silences an alarm raised in CI.



The well-known techniques of demand-driven analysis and incremental analysis help address this challenge.
De\-mand-driven analyses compute only those results needed to answer a set of extrinsicially-provided \emph{queries}, while incremental analyses speed up re-analysis of an \emph{edited} program by re-using as many previously-computed results as possible.
Powerful frameworks for incremental (e.g., \cite{DBLP:conf/kbse/SzaboEV16,DBLP:conf/icse/ArztB14}) \emph{or} demand-driven (e.g., \cite{DBLP:conf/sigsoft/HorwitzRS95}) static analysis do exist, but nearly all such frameworks target restricted analysis domains (e.g., finite or finite-height domains), whereas well-known analyses like octagon and shape analysis require an infinite-height abstract domain.
There are also approaches to adapt summary-based analyses to offer coarse-grained method- or file-level incrementality (e.g., \cite{DBLP:conf/nfm/CalcagnoD11,DBLP:journals/cacm/DistefanoFLO19, DBLP:conf/foveoos/FahndrichL10}). Though these approaches effectively scale to industrial codebases in CI pipelines, they are not intended to achieve real-time interactivity during the development process.


In contrast, our aim is to support fine-grained incremental {\em and} demand-driven analysis over \emph{arbitrary} abstract domains expressed in general-purpose languages,
thus enabling the reuse of existing optimized abstract domain implementations at interactive speeds.
To our best knowledge, no general technique exists to automatically compute a demand-driven or incremental version of an arbitrary abstract interpretation with arbitrary widening operators, a key requirement to ensure termination in more complex analyses.

This paper presents \emph{demanded abstract interpretation}, an abstract interpretation framework with first-class support for {\em both} demand-driven and incremental analysis.
We build on abstract interpretation~\cite{DBLP:conf/popl/CousotC77}, which provides a methodology for expressing static analyses and guaranteeing their correctness, and take inspiration from work on general incremental computation using ``demanded computation graphs''~\cite{DBLP:conf/pldi/HammerKHF14}.

Our framework reifies the abstract interpretation (AI) of a program as a dynamically evolving \emph{demanded abstract interpretation graph} (DAIG),
which explicitly represents program statements, abstract states, and the dependency structure of analysis computations.
In this representation, program edits, client-issued queries, and the evaluation of abstract semantics can all be treated uniformly.
Cyclic control flow is a key difficulty for this approach,
since cyclic dependencies lead to unclear evaluation semantics.
We define an operational semantics for DAIGs that preserves an acyclic invariant while modifying and extending the graph on demand,
thus soundly analyzing loops with guaranteed termination, assuming termination of the underlying abstract interpretation.

The DAIG encoding and evaluation enables efficient abstract interpretation in an {\em interactive} mode, analyzing a minimal number of statements to respond to queries with maximal reuse of previously-computed results.
In particular, this paper makes the following key contributions:
\begin{itemize}[leftmargin=1em]
\item We introduce a framework for {\em demanded abstract interpretation} in which program syntax and analysis computation structure are reified into {\em demanded abstract interpretation graphs}~(DAIGs)~(\cref{sec:daigs}).
\item We specify an operational semantics for DAIGs that realizes incremental updates and demand-driven evaluation via demanded unrolling of abstract interpretation fixed-point computations~(\cref{sec:daig_semantics}).
\item We prove that demanded abstract interpretation preserves soundness and termination, and that its results are from-scratch consistent with classical abstract interpretation by global fixed-point iteration~(\cref{sec:metatheory}).
\item We provide evidence for the expressivity and efficacy of demanded abstract interpretation using a prototype framework instantiated with interval, octagon, and shape domains~(\cref{sec:eval}). In our experiments, DAIGs support context-sensitive interprocedural analysis at interactive speeds, answering 95\% of queries within \demandedpninetyfive~seconds.
\end{itemize}

\section{Overview}\label{sec:overview}
\noindent
\cref{fig:overview:ast} shows a simple imperative program that appends two linked lists.
Given well-formed (i.e., null-terminated and acyclic) input lists \lstinline{p} and \lstinline{q}, \lstinline{append} must return a well-formed list and not dereference \lstinline{null} in order to be correct.
These properties can be verified using a separation logic, abstract interpretation-based shape analysis~\cite{DBLP:conf/fmco/BerdineCO05,DBLP:conf/tacas/DistefanoOY06,magill2006inferring}, tracking facts like $\code{lseg}(p,null)$ to represent well-formedness of list \lstinline{p}.
Our goal is to enable interactive performance for arbitrary abstract interpretations, including such analyses, in response to a user's edits and queries.

In this section, we illustrate our approach to \emph{demanded abstract interpretation} by example.
We demonstrate how abstract interpretation of forwards control flow is reified in a DAIG (\cref{overview:cfgtodaig}), and then show how the DAIG supports both demand-driven and incremental interactions with the underlying abstract interpretation (\cref{overview:demand-and-incremental}).
Finally, we highlight key difficulties introduced by cyclic control flow, show how analysis thereof can be encoded acyclically, and demonstrate how our operational semantics evolve the DAIG on-demand to soundly compute fixed points (\cref{overview:cycles}).

\subsection{Reifying Abstract Interpretation in DAIGs}
\label{overview:cfgtodaig}

The \code{append} procedure from \cref{fig:overview:ast} may equivalently be represented as a control-flow graph (CFG) as shown in \cref{fig:overview:cfg}, with vertices for program locations and edges labelled by atomic program statements.\footnote{As is standard, we have simply broken down the guard conditions from \lstinline{if} and \lstinline{while} into \lstinline{assume} guard statements for each side of a branch.} A classical abstract interpreter analyzes such a program by starting with some initial abstract state at the entry location and 
applying an abstract transfer function $\denote\cdot^\sharp$ to interpret statements,
a join operator $\sqcup$ at nodes with multiple predecessors, and a widen operator $\nabla$ at cycles as needed until a fixed point is reached.

\begin{figure}\rule{1.5em}{0pt}\lststopn
\begin{lstlisting}[language=jsimp,style=number] 
function append(p: List, q: List): List { $\makeatletter\global\lst@savelstnumber=-1\lststartn\makeatother$
  if (p == null) {
    return q; }
  var r: List = p;
  while (r.next != null) {
    r = r.next; }
  r.next = q;
  return p;$\gdef\thelstnumber{\lstnumprefix{\text{ret}}}$
}
\end{lstlisting}
\vspace{-1em}
\caption{A procedure to append two linked lists. The labels $\ell_i$ mark program locations in its control flow.}
\label{fig:overview:ast}
\vspace{-1em}
\end{figure}

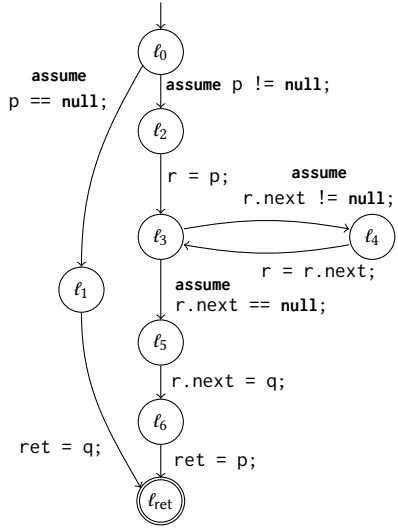
\begin{figure}
  \begin{tikzpicture}[align=left,font=\footnotesize]
  \tikzset{every node/.style={minimum width=0.6cm, inner sep = 0em}}
  
  \node (l0) [circle,draw] {$\ell_0$};
  \node (l2) [circle, draw, node distance=3em, below of =l0] {$\ell_2$};
  \node (l3) [circle, draw, node distance=4em, below of =l2] {$\ell_3$};
  \node (l5) [circle, draw, node distance=4em, below of =l3] {$\ell_5$};
  \node (l6) [circle, draw, node distance=3em, below of =l5] {$\ell_6$};
  \node (lret) [circle, draw, double, node distance=3em, below of =l6] {$\ell_{\tiny\textsf{ret}}$};

  \node (l1) [circle, draw, left of =l3, node distance = 3em,yshift=-2em] {$\ell_1$};
  \node (l4) [circle, draw, right of =l3, node distance = 8em] {$\ell_4$};

  \draw[->] ([yshift=1em]l0.north) to (l0.north) node [left of = l0, align = center,xshift=-1em, yshift=-1.5em] {\usebox{\SBoxAssume}\\ \texttt{p == \usebox{\SBoxNull};}};
  \draw[->] (l0.south) to (l2.north) node [right of=l0, align=left,yshift=-1.25em,xshift=0.5em] {\texttt{\usebox{\SBoxAssume} p != \usebox{\SBoxNull};}};
  \draw[->] (l2.south) to  (l3.north) node [right of = l2, align=left,xshift=-1.4em,yshift=-1.8em] {\texttt{r = p;}};
  \draw[->] (l3.south) to (l5.north) node [align = left, right of = l3, xshift=0.6em, yshift=-2.3em] {\usebox{\SBoxAssume}\\\vspace{-1.5em}\\\texttt{r.next == \usebox{\SBoxNull};}};
  \draw[->] (l5.south) to (l6.north) node [align = left, right of =l5, yshift=-1.5em,xshift=-0.25em] {\texttt{r.next = q;}};
  \draw[->] (l6.south) to (lret.north) node [align = left, right of =l6, yshift=-1.5em,xshift=-0.75em] {\texttt{ret = p;}};

  \draw[->] ([yshift=0.3em]l3.east) to[out=10,in=170] ([yshift=0.3em]l4.west) node [align=center, above of=l4,xshift=-2em,yshift=-1em,black]{\usebox{\SBoxAssume}\\ \texttt{r.next != \usebox{\SBoxNull};}};
  \draw[->] ([yshift=-0.3em]l4.west) to[out=190,in=350] ([yshift=-0.3em]l3.east) node [below of=l4,yshift=1.5em,xshift=-2em,black]{\texttt{r = r.next;}};

  \draw[->]([yshift=-0.5em,xshift=0.2em]l0.west) to[out=240,in=90] (l1.north);
  \draw[->](l1.south) to[out=270,in=120] ([yshift=0.5em,xshift=0.12em]lret.west) node [xshift=-3em,yshift=1.5em] {\texttt{ret = q;}};
  
%
%
%
\end{tikzpicture}
  \caption{%
 The control-flow graph (CFG) of the \code{append} procedure from \cref{fig:overview:ast}.
    }
  \label{fig:overview:cfg}
  \vspace{-1em}
\end{figure}

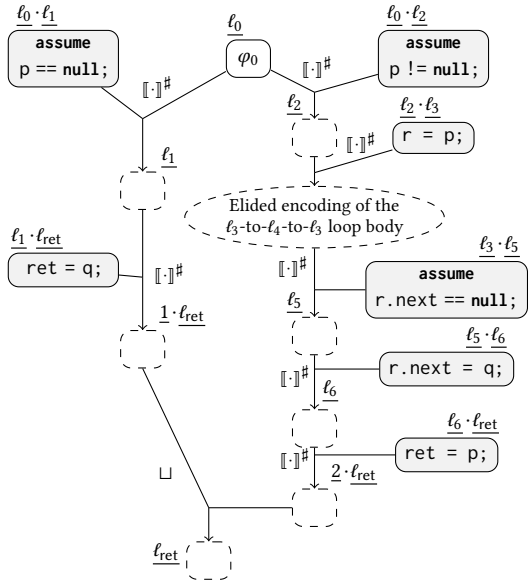
\begin{figure}
  \begin{tikzpicture}[font=\footnotesize,align=center]
  
  \node (s0) [rectangle, text width=3.5em, draw, fill=black!5!white, label={[yshift=-0.25em,xshift=-1em]$\underline{\ell_0}\nameprod\underline{\ell_1}$},rounded corners] {\texttt{\usebox{\SBoxAssume} p\;==\;\usebox{\SBoxNull};}};
  \node (phi0) [rectangle, text width=1em, text height = 0.75em, draw, rounded corners, label={[yshift=-0.25em,xshift=-0.5em]$\underline{\ell_0}$}, right of=s0, node distance = 7em,text depth=0.25em,text height = 0.5em] {$\varphi_0$};
  \node (s1) [rectangle, text width =3.5em, draw, fill=black!5!white,rounded corners, label={[yshift=-0.25em,xshift=-1em]$\underline{\ell_0}\nameprod\underline{\ell_2}$}, right of=phi0, node distance = 7em] {\texttt{\usebox{\SBoxAssume} p\;!=\;\usebox{\SBoxNull};}};

  \node (phi1) [rectangle, draw, dashed, align=left, text width=1em, text height = 0.75em, rounded corners, label={[yshift=-0.25em,xshift=1em]$\underline{\ell_1}$}, below of=phi0, node distance = 5em,xshift=-4em] {};
  
  \node (s2) [rectangle, draw, label={[yshift=-0.25em,xshift=-0.5em]$\underline{\ell_2}\nameprod\underline{\ell_3}$}, below of=phi0, node distance= 3em, xshift = 7em, fill=black!5!white,rounded corners] {\texttt{r = p;}};
  \node (phi2) [rectangle, draw, dashed, text width=1em, text height = 0.75em,rounded corners, label={[yshift=-0.25em,xshift=-0.75em]$\underline{\ell_2}$}, below of=phi0, node distance = 3em, xshift=2.5em] {};
  
  \node (s8) [rectangle, draw, text width=3.5em, label={[yshift=-0.25em,xshift=-1em]$\underline{\ell_1}\nameprod\underline{\ell_{\tiny\textsf{ret}}}$}, below of=phi0, node distance= 8em, xshift = -7em, fill=black!5!white,rounded corners] {\texttt{ret = q;}};

  \node (elided_loop) [ellipse, draw, align=center, dashed, below of = phi2, inner sep=1pt, node distance = 3em,font=\scriptsize] {Elided encoding of the\\ $\ell_3$-to-$\ell_4$-to-$\ell_3$ loop body};
  
  \node (phi8) [rectangle, draw, dashed, text width=1em, text height = 0.75em,rounded corners, label={[yshift=-0.25em,xshift=1.5em]$\underline 1 \nameprod\underline{\ell_{\tiny\textsf{ret}}}$}, below of=phi1, node distance = 6em] {};

  \node (s5) [rectangle, draw, label={[yshift=-0.25em,xshift=2em]$\underline{\ell_3}\nameprod\underline{\ell_5}$}, below of=elided_loop, node distance=2.75em, xshift = 5em, fill=black!5!white,rounded corners] {\usebox{\SBoxAssume} \\\texttt{r.next\;==\;\usebox{\SBoxNull};}};

  \node (phi5) [rectangle, draw, dashed, text width=1em, text height = 0.75em,rounded corners, label={[yshift=-0.25em,xshift=-0.75em]$\underline{\ell_5}$}, below of=elided_loop, node distance = 4.5em] {};

    \node (s6) [rectangle, draw, label={[yshift=-0.25em,xshift=1.5em]$\underline{\ell_5}\nameprod\underline{\ell_6}$}, below of=s5, node distance=3em, fill=black!5!white,rounded corners] {\texttt{r.next = q;}};
  \node (phi6) [rectangle, draw, dashed, text width=1em, text height = 0.75em,rounded corners, label={[yshift=-0.1em,xshift=0.6em]$\underline{\ell_6}$}, below of=phi5, node distance = 3.5em, ] {};

  \node (s7) [rectangle, draw, label={[yshift=-0.25em,xshift=1em]$\underline{\ell_6}\nameprod\underline{\ell_{\tiny\textsf{ret}}}$}, below of=s6, node distance=3.25em,  fill=black!5!white,rounded corners] {\texttt{ret = p;}};
  \node (phi7) [rectangle, draw, dashed, text width=1em, text height = 0.75em,rounded corners, label={[yshift=-0.25em,xshift=1.5em]$\underline 2 \nameprod\underline{\ell_{\tiny\textsf{ret}}}$}, below of=phi6, node distance = 3em, ] {};

  \node (phi_ret) [rectangle, draw, dashed, text width=1em, text height = 0.75em,rounded corners, label={[left,xshift=-0.75em,yshift=-0.5em]$\underline{\ell_{\tiny\textsf{ret}}}$}, below of=phi7, node distance = 2em, xshift=-4em ] {};
\draw [-] (s1) -- ([yshift=1em]phi2.north);
\draw [-] (phi0) -- ([yshift=1em]phi2.north) node [near end,yshift=0.75em,xshift=0.5em, font=\tiny] {$\denote\cdot^\sharp$};
\draw [->] ([yshift=1em,]phi2.north) -- (phi2.north);

\draw [-] (s0) -- ([yshift=2em]phi1.north);
\draw [-] (phi0) -- ([yshift=2em]phi1.north) node [near end, xshift=-0.25em,yshift=0.75em,font=\tiny] {$\denote\cdot^\sharp$};
\draw [->] ([yshift=2em]phi1.north) -- (phi1.north);

\draw [-] (s8) -- ([yshift=2em]phi8.north);
\draw [-] (phi1) -- ([yshift=2em]phi8.north) node [near end, xshift=1em,yshift=-0.5em,font=\tiny] {$\denote\cdot^\sharp$};
\draw [->] ([yshift=2em]phi8.north) -- (phi8);

\draw [-] (s2) -- ([yshift=0.5em]elided_loop.north);
\draw [-] (phi2.south) -- ([yshift=0.5em]elided_loop.north) node [near end, xshift=1.75em,yshift=1em,font=\tiny] {$\denote\cdot^\sharp$};
\draw [->] ([yshift=0.5em]elided_loop.north) -- (elided_loop.north);

\draw [-] (s5.west) -- (phi5.north|-s5.west);
\draw [-] (elided_loop) -- (phi5.north|-s5.west) node [near end, xshift=-0.75em,yshift=0.5em,font=\tiny] {$\denote\cdot^\sharp$ };
\draw [->] (phi5.north |- s5.west) -- (phi5.north);

\draw [-] (s6) -- (phi6.north|-s6);
\draw [->] (phi5) -- (phi6.north) node [near end, xshift=-0.75em,yshift=0.75em,font=\tiny] {$\denote\cdot^\sharp$};

\draw [-] (s7) -- (phi7.north|-s7);
\draw [->] (phi6.south) -- (phi7.north) node [near end, xshift=-0.75em,yshift=0.75em,font=\tiny] {$\denote\cdot^\sharp$};

\draw [-] (phi7.west) -- (phi_ret.north|-phi7.west);
\draw [-] (phi8.south) -- (phi_ret.north|-phi7.west) node [near end, xshift=-1em] {$\sqcup$};
\draw [->] (phi_ret.north|-phi7.west) -- (phi_ret.north);

\end{tikzpicture}
  \caption{A demanded abstract interpretation graph (DAIG) for the program given in \cref{fig:overview:ast} before any queries are issued. The elided loop encoding is shown in \cref{fig:overview:daig_unroll}.}
  \label{fig:overview:daig}
  \vspace{-0.5em}
\end{figure}


The demanded abstract interpretation graph (DAIG) shown in \cref{fig:overview:daig} reifies the computational structure of such an abstract interpretation of the \cref{fig:overview:cfg} CFG.
Its vertices are u\-ni\-que\-ly-named mutable reference cells containing program syntax or abstract state, and its edges
fully specify the computations of an abstract interpretation.
Names identify values for reuse across edits and queries, and hence must {\em uniquely} identify the inputs and intermediate results of the abstract interpretation.
In \cref{fig:overview:daig} and throughout this paper, {\em underlined} symbols denote a name derived from that symbol: hashes, essentially.

\newsavebox{\SBoxAssumePIsNull}\sbox{\SBoxAssumePIsNull}{\lstinline{assume(p == null)}}
%

To encode abstract interpretation computations, DAIG edges are labelled by a symbol for an abstract interpretation function and connect cells storing the function inputs to the cell storing the output, capturing the dependency structure of the analysis computation.%
\footnote{More precisely, DAIGs have {\em hyper-}edges, since they connect multiple sources (function inputs) to one destination (function output).}
For example, the computation of the abstract transfer function over the CFG edge $\ell_0 \rightarrow \ell_1$ is encoded in \cref{fig:overview:daig} as a DAIG edge with input cells $\underline{\ell_0}$ and $\underline{\ell_0}\cdot\underline{\ell_1}$ (respectively containing the fixed-point state at $\ell_0$ and the corresponding statement $s_0\colon \usebox{\SBoxAssumePIsNull}$), labelled by the abstract transfer function symbol~$\denote\cdot^\sharp$.

\newsavebox{\SBoxRetGetsQ}\sbox{\SBoxRetGetsQ}{\lstinline{ret = q}}
\subsection{Demand-Driven and Incremental Analysis}\label{overview:demand-and-incremental}

Next, we demonstrate how a DAIG encoding naturally supports demand-driven and incremental analysis.
%
We use the aformentioned shape-analysis domain for our example, a separation logic-based domain with a ``list segment'' primitive $\code{lseg}(x,y)$ that abstracts the heaplet containing a list segment from $x$ to $y$.\footnote{That is, a sequence of iterated \code{\footnotesize next} pointer dereferences from $x$ to $y$.}  This domain is of infinite height, absent a best abstraction function, and with complex widening operators, and therefore incompatible with previous frameworks that restrict the domain form.
%

\newsavebox{\SBoxDaigDemandFig}
\begin{lrbox}{\SBoxDaigDemandFig}
\begin{tikzpicture}[font=\footnotesize,align=center]
  
  \node (s0) [rectangle, rounded corners,text width=3.5em, draw, fill=black!5!white, label={[yshift=-0.25em,xshift=-1em]$\underline{\ell_0}\nameprod\underline{\ell_1}$}] {\usebox{\SBoxAssume}\\\texttt{p == \usebox{\SBoxNull};}};
  \node (phi0) [rectangle, text width=6em, draw, rounded corners, label={[yshift=-0.25em,xshift=-1.5em]$\underline{\ell_0}$}, right of=s0, node distance = 6em] {$\varphi_0\colon\begin{array}{@{}l@{}}\texttt{lseg}(\texttt q,\textit{null})\,\ast\\\texttt{lseg}(\texttt p,\textit{null})\end{array}$};

  \node (phi1) [rectangle, draw, align=left, text width=4em,rounded corners, label={[yshift=-0.25em,xshift=1.5em]$\underline{\ell_1}$}, below of=phi0, node distance = 5em, xshift=-1em] {$\begin{array}{@{}l@{}}\texttt{lseg}(\texttt q,\textit{null})\\\wedge~\texttt p=\textit{null}\end{array}$};
  \node (s8) [rectangle, draw, text width=3.5em, label={[yshift=-0.25em,xshift=-1em]$\underline{\ell_1}\nameprod\underline{\ell_{\tiny\textsf{ret}}}$}, below of=phi0, node distance= 4em, xshift = -6em, rounded corners, fill=black!5!white] {\texttt{ret = q;}};

  \node (phi8) [rectangle, draw, text width=5em,rounded corners, label={[yshift=-0.25em,xshift=1.5em]$\underline 1 \nameprod\underline{\ell_{\tiny\textsf{ret}}}$}, below of=phi1, node distance = 4.5em, xshift=-3em] {$\begin{array}{@{}l@{}}\texttt{lseg}(\texttt {ret},\textit{null})\\\wedge~\texttt p=\textit{null}\end{array}$};

\draw [-] (s0) -- ([yshift=1.5em,xshift=-1em]phi1.north);
\draw [-] (phi0) -- ([yshift=1.5em,xshift=-1em]phi1.north) node [near end, xshift=-1.3em,yshift=-0.5em,font=\tiny] {$\denote\cdot^\sharp$};
\draw [->] ([yshift=1.5em,xshift=-1em]phi1.north) -- ([xshift=-1em]phi1.north);

\draw [-] (s8) -- ([yshift=1.5em,xshift=-1em]phi8.north);
\draw [-] (phi1) -- ([yshift=1.5em,xshift=-1em]phi8.north) node [near end, xshift=-1.3em,yshift=-0.5em,font=\tiny] {$\denote\cdot^\sharp$};
\draw [->] ([yshift=1.5em,xshift=-1em]phi8.north) -- ([xshift=-1em]phi8.north);
\end{tikzpicture}
\end{lrbox}
\newsavebox{\SBoxDaigEditFig}
\begin{lrbox}{\SBoxDaigEditFig}
\begin{tikzpicture}[black!60!white,font=\footnotesize,align=left]

  \node (l1) [rectangle, draw, align=center, text width=1em, text height=0.5em,rounded corners, label={[yshift=-0.25em,xshift=-0.5em]$\underline{\ell_1}$}] {\makebox[0.75em]{$\varphi$}};
  \node (l1l7) [rectangle, draw=black!50!green, text=black!50!green, text width=4em, label={[yshift=-0.25em,xshift=-1em,color=black!50!green]$\underline{\ell_1}\nameprod\underline{\ell_7}$}, left of=l1, node distance= 5em, fill=black!5!white, rounded corners] {\texttt{print("p}\\\texttt{is null");}};

  \node (l7) [rectangle, draw, dashed, color=black!50!green, align=left, text width=1em, text height=0.75em,rounded corners, label={[yshift=-0.25em,xshift=-0.5em,color=black!50!green]$\underline{\ell_7}$}, below of = l1, node distance = 4em,xshift=-2em] {};

  \node (1lret) [rectangle, align=center, draw, dashed, black!50!red, text width=1em,text height =0.75em,rounded corners, below of=l7,node distance=4em,label={[black!50!red, yshift=-0.25em,xshift=-1em]$\underline 1 \nameprod\underline{\ell_{\tiny\textsf{ret}}}$}] {\makebox[0.75em]{\sout{$\varphi'$}}};
  \node (2lret) [rectangle, draw, text width=1em, text height = 0.75em,rounded corners, right of = 1lret, node distance=4em,label={[yshift=-0.25em,xshift=-0.7em]$\underline 2 \nameprod\underline{\ell_{\tiny\textsf{ret}}}$}] {};
  \node (lret) [rectangle, draw, dashed, align=center, black!50!red, text width=1em, text height = 0.75em,rounded corners, below of=2lret, xshift=-2em, label={[black!50!red,left,xshift=-0.75em,yshift=-0.5em]$\underline{\ell_{\tiny\textsf{ret}}}$}] {\makebox[1em]{\sout{$\varphi''$}}};
  
  \node (l7lret) [rectangle, draw, text width=3.5em, label={[yshift=-0.25em,xshift=-1em]$\underline{\ell_7}\nameprod\underline{\ell_{\tiny\textsf{ret}}}$}, left of=l7, node distance= 4em,  fill=black!5!white,rounded corners] {\texttt{ret = p;}};

  \node[above of=l1,xshift=0.25em] {$\vdots$};
  \draw[->] ([yshift=1em,xshift=0.25em]l1.north) -- ([xshift=0.25em]l1.north);

  \node[above of=2lret,xshift=0.45em] {$\vdots$};
  \draw[->] ([yshift=1em,xshift=0.45em]2lret.north) -- ([xshift=0.45em]2lret.north);

  \node (prel7) [black!50!green, above of=l7, node distance =2em,coordinate,xshift=0.3em,label={[right,black!50!green,xshift=0.2em]$\denote\cdot^\sharp$}] {};
  \draw[-,black!50!green] (l1.south) -- (prel7);
  \draw[-,black!50!green] (l1l7) -- (prel7);
  \draw[->,black!50!green] (prel7) -- ([xshift=0.3em]l7.north);

  \node (pre1lret) [black!50!red, above of=1lret, node distance =2em,coordinate,xshift=0.3em,label={[right,black!50!red]$\denote\cdot^\sharp$}] {};
  \draw[-,black!50!red] ([xshift=0.3em]l7.south) -- (pre1lret);
  \draw[-,black!50!red] (l7lret) -- (pre1lret);
  \draw[->,black!50!red] (pre1lret) -- ([xshift=0.3em]1lret.north);

  \node (prelret) [black!50!red, above of=lret, node distance =2em,coordinate,label={[above,black!50!red]$\sqcup$}] {};
  \draw[-,black!50!red] (1lret) -- (prelret);
  \draw[-,black!50!red] (2lret) -- (prelret);
  \draw[->,black!50!red] (prelret) -- (lret);

\end{tikzpicture}
\end{lrbox}
\newsavebox{\SBoxDaigUnrollFig}
\begin{lrbox}{\SBoxDaigUnrollFig}
\begin{tikzpicture}[font=\footnotesize,align=center, daig node/.style={rectangle, draw, rounded corners, text width=0.75em, text height = 0.75em}, darkred/.style={color=black!40!red}]
\newcommand{\iterloc}[2]{$\underline{\ell_{#1}}^{\hspace{0.05em}(#2)}$}

\node(l30)[daig node, label={[left,yshift=0.3em,xshift=0.4em]\iterloc{3}{0}}] {};

\node(l3l4)[rectangle,draw,rounded corners,fill=black!5!white,label={[yshift=-0.25em]$\underline{\ell_3}\nameprod\underline{\ell_4}$}, left of = l30, node distance = 6em] {\usebox{\SBoxAssume}\\\texttt{r.next != \usebox{\SBoxNull};}};

\node(l4l3)[rectangle,draw,rounded corners, fill=black!5!white,label={[yshift=-0.25em]$\underline{\ell_4}\nameprod\underline{\ell_3}$}, below of = l3l4, node distance = 4em] {\texttt{r = r.next;}};

\node(l40)[daig node, label={[left,yshift=0.3em,xshift=0.2em]\iterloc{4}{0}}, below of = l30, node distance = 4em] {};

\node(l30l31)[daig node, label={[left,yshift=0.6em,xshift=0.6em]\iterloc{3}{0}\hspace{-0.2em}$\nameprod$\hspace{0.2em}\iterloc{3}{1}}, right of = l40, node distance = 4em] {};

\node(l31)[daig node, label={[left,yshift=0.3em,xshift=0.4em]\iterloc{3}{1}}, right of = l30, node distance = 8em] {};

\node(l41)[darkred, dashed, daig node, label={[darkred, left,yshift=0.3em,xshift=0.2em]\iterloc{4}{1}}, below of = l31, node distance = 4em] {};

\node(l31l32)[darkred, dashed, daig node, label={[darkred,left,yshift=0.6em,xshift=0.6em]\iterloc{3}{1}\hspace{-0.2em}$\nameprod$\hspace{0.2em}\iterloc{3}{2}}, right of = l41, node distance = 4em] {};

\node(l32)[darkred, dashed, daig node, label={[darkred, left,yshift=0.3em,xshift=0.4em]\iterloc{3}{2}}, right of = l31, node distance = 8em] {};

\node(l3)[daig node, dashed, label={[yshift=-0.2em,xshift=0.5em]$\underline{\ell_3}$}, right of = l32, node distance = 6em] {};

\node[above of=l30,xshift=0.25em] {$\vdots$};
\draw[->] ([yshift=1em,xshift=0.25em]l30.north) -- ([xshift=0.25em]l30.north);
\node[below of=l3,yshift=0.7em] {$\vdots$};
\draw[->] (l3.south) -- ([yshift=-1em]l3.south);

\node (prel40) [above of=l40, node distance = 2.5em,coordinate,label={[left,xshift=-0.6em,yshift=-0.2em,font=\tiny]$\denote{\cdot}^\sharp$}] {};
\draw[->] (prel40) -- (l40);
\draw[-] (l3l4) -- (prel40);
\draw[-] (l30) -- (prel40);

\draw[->] (l40.south) to[out=330,in=230] ([xshift=-0.5em,yshift=0.1em]l30l31.south);
\draw[-] (l4l3) to[out=350,in=230] ([xshift=-0.5em,yshift=0.1em]l30l31.south) node[right of = l40,font=\tiny,xshift=-0.5em,yshift=-0.6em]{$\denote{\cdot}^\sharp$};

\draw[->](l30.east) -- (l31.west);
\draw[-] ([xshift=0.25em]l30l31.north) to[out=90,in=180] ([xshift=-1.5em]l31.west) node [left of = l31,yshift=0.5em]{$\nabla$};

\node (prel41) [above of=l41, node distance = 2.5em,coordinate,label={[darkred,left,xshift=-0.6em,yshift=-0.4em,font=\tiny]$\denote{\cdot}^\sharp$}] {};
\draw[->,darkred] (prel41) -- (l41);
\draw[-,darkred] ([yshift=-0.75em]l3l4.east) -- (prel41);
\draw[-,darkred] (l31) -- (prel41);

\draw[->,darkred] (l41.south) to[out=330,in=210] ([xshift=-0.55em,yshift=0.1em]l31l32.south);
\draw[-,darkred] (l4l3) to[out=345,in=210] ([xshift=-0.55em,yshift=0.1em]l31l32.south) node[darkred,right of = l41,font=\tiny,xshift=-0.5em,yshift=-0.4em]{$\denote{\cdot}^\sharp$};

\draw[->,darkred] (l31.east) -- (l32.west);
\draw[-,darkred] ([xshift=0.25em]l31l32.north) to[out=90,in=180] ([xshift=-1.5em]l32.west) node [darkred,left of = l32,yshift=0.5em]{$\nabla$};

\draw[->,darkred] (l31.north) to[out=20,in=160] ([xshift=-0.55em,yshift=-0.1em]l3.north);
\draw[-,darkred] (l32.north) to[out=20,in=160] ([xshift=-0.55em,yshift=-0.1em]l3.north);

\draw[->,color=black!50!white,dotted] ([xshift=0.5em]l30.north) to[out=20,in=160] ([xshift=-0.55em,yshift=-0.1em]l3.north) node[above of =l32,darkred,xshift=3.5em,yshift=-1em] {\textsf{fix}};
\draw[-,color=black!50!white,dotted] (l31.north) to[out=40,in=180] ([xshift=-11em,yshift=2em]l3.north)node[above of=l31,color=black!50!white,xshift=1em,yshift=0.5em] {\textsf{fix}};

\end{tikzpicture}
\end{lrbox}
\newcommand{\daigfigbox}[2]{\subcaptionbox{#2}{\scalebox{0.87}{\usebox{#1}}}}
\begin{figure*}
\daigfigbox{\SBoxDaigDemandFig}{Demand-driven query\label{fig:overview:daig_demand}}
\hfill
\daigfigbox{\SBoxDaigEditFig}{Incremental edit\label{fig:overview:daig_edit}}
\hfill
\daigfigbox{\SBoxDaigUnrollFig}{Cyclic-control flow and demanded fixed points\label{fig:overview:daig_unroll}}
\caption{%
Demanded abstract interpretation:
(\subref{fig:overview:daig_demand})
Demanding a value for $\underline 1\nameprod\underline{\ell_{\tiny\textsf{ret}}}$
recursively triggers demand for its dependencies and is resolved by computing its value from the statements in $\underline{\ell_0}\nameprod\underline{\ell_1}$ and $\underline{\ell_1}\nameprod\underline{\ell_{\tiny\textsf{ret}}}$ and the initial state $\varphi_0$ in $\underline{\ell_0}$.
We show only the relevant subgraph here, but this operation occurs in the full DAIG of~\cref{fig:overview:daig}.
(\subref{fig:overview:daig_edit})
DAIG from \cref{fig:overview:daig} updated to reflect nodes added ( $\underline{\ell_1}\nameprod\underline{\ell_7}$ and $\underline {\ell_7}$) and potentially affected ($\underline 1 \nameprod \underline {\ell_{\textsf{ret}}}$ and $\underline {\ell_{\textsf{ret}}}$) by the edit in \cref{overview:demand-and-incremental}.  All other nodes are unchanged.
(\subref{fig:overview:daig_unroll})
DAIG for the $\ell_3$-to-$\ell_4$-to-$\ell_3$ loop of \cref{fig:overview:cfg} after one demanded unrolling, with the new DAIG region shown in {\color{black!40!red}red}  and the (removed) pre-unrolling \textsf{fix} edge shown in dotted {\color{black!50!white} grey}.  Note that cells containing program syntax are not duplicated.
The DAIG with the black vertices and edges along with the {\color{black!50!white} grey} edge (but not the {\color{black!40!red}red} ones) is the initial sub-DAIG for the dashed ellipse in \cref{fig:overview:daig}.
}
\end{figure*}

In \cref{fig:overview:daig_demand}, we show the result of evaluating a demand query on our example DAIG.
Suppose a client issues a query for the $\underline 1\nameprod\underline{\ell_{\tiny\textsf{ret}}}$ cell in \cref{fig:overview:daig}, the abstract state corresponding to the \lstinline{return q} statement at $\ell_1$ in \cref{fig:overview:ast}.
Since the $\underline 1\nameprod\underline{\ell_{\tiny\textsf{ret}}}$ cell has predecessors $\underline{\ell_1}\nameprod\underline{\ell_{\tiny\textsf{ret}}}$ and $\underline{\ell_1}$, we issue requests for the values of those cells.  
Cell $\underline{\ell_1}$ is empty, but depends on $\underline{\ell_0}\nameprod\underline{\ell_1}$ and $\underline{\ell_0}$, so more requests are issued.
Both of those cells hold values, 
so we can compute and store the value of $\underline{\ell_1}$. 
Now, having satisfied its dependencies, we can compute the value of $\underline 1\cdot\underline{\ell_{\tiny\textsf{ret}}}$, as shown in \cref{fig:overview:daig_demand}. 
Note that DAIGs are always acyclic, so this recursive traversal of dependencies is well-founded.

Crucially, these results are now {\em memoized} for future incremental reuse; a subsequent query for $\underline{\ell_\text{\tiny ret}}$, for example, will memo match on $\underline 1\nameprod\underline{\ell_{\tiny\textsf{ret}}}$ and only need to compute $\underline 2 \nameprod \underline{\ell_{\tiny\textsf{ret}}}$ and its dependencies from scratch.
This fine-grained reuse of intermediate abstract interpretation results is a key feature of the DAIG encoding for demand-driven analysis.
To handle developer edits to code, DAIGs are also naturally {\em incremental}, efficiently recomputing and reusing analysis results across multiple program versions, following the incremental computation with names approach~\cite{DBLP:conf/oopsla/HammerDHLFHH15}.

Consider a program edit which adds a logging statement \code{print("p is null")} just before the \lstinline{return} at $\ell_1$ in \cref{fig:overview:ast}.
Intuitively, program behaviors are unchanged at those locations unreachable from the added statement, so an incremental analysis should only need to re-analyze the sub-DAIG reachable from the new statement.

\cref{fig:overview:daig_edit} illustrates this program edit's effect on the DAIG.  The \textcolor{black!60!green}{green} nodes correspond to the added statement cell $\underline{\ell_1}\nameprod\underline{\ell_7}$ and its corresponding abstract state cell $\underline{\ell_7}$.
Nodes forward-reachable from the green nodes --- those marked in \textcolor{black!50!red}{red} --- are invalidated (a.k.a. ``dirtied'') by our incremental computation engine. In particular, cells
$\underline{\ell_1}\nameprod\underline{\ell_{\tiny\textsf{ret}}}$ and
$\underline{\ell_{\tiny\textsf{ret}}}$ containing abstract states $\varphi'$ and $\varphi''$, respectively, are dirtied.

Crucially, while nodes are dirtied {\em eagerly}, they are recomputed from up-to-date inputs {\em lazily}, only when demanded.
That is, the DAIG encoding allows our analysis to avoid constant recomputation of an analysis as a program is edited, instead computing results on demand while soundly keeping track of which intermediate results --- possibly from a previous program version --- are available for reuse.
For example, assuming that $\underline{\ell_1}$ and $\underline{2}\nameprod\underline{\ell_{\tiny\textsf{ret}}}$ were both computed before the edit, a query for $\underline{\ell_\text{\tiny ret}}$ must execute only two transfers and one join: the \textcolor{black!50!red}{red} and \textcolor{black!60!green}{green} edges of \cref{fig:overview:daig_edit}.
This represents a significant savings over recomputing the entire analysis --- including the loop fixed point --- as would be necessary without incremental analysis.

If desired, an auxiliary memoization (memo) table can also be used to cache computations independent of program locations to enable further incrementalization, with names based on the input values (e.g., memoizing $\denote{s_0}^\sharp(\varphi_0)$ in
a cell named $\underline{\denote\cdot^\sharp}\cdot\underline{s_0}\cdot\underline{\varphi_0}$). 
As with batch analysis, it is sound to drop cached results from the DAIG and/or memo table and later recompute those results if needed, trading efficiency of reuse for a lower memory footprint.


\subsection{Cyclic Control Flow and Demanded Fixed Points}\label{overview:cycles}
As shown in \cref{overview:cfgtodaig}, encoding program structure and analysis data-flow into a dependency graph is relatively straightforward when the control-flow graph is acyclic.
However, when handling loops or recursion, like lines $\ell_3$ and $\ell_4$ in \cref{fig:overview:ast}, an abstract interpreter's fixed-point computation is inherently cyclic.
Properly handling these cyclic control-flow and data-flow dependency structures is the crux of
realizing demanded abstract interpretation.
For instance, introducing cyclic dependencies in the DAIG yields an unclear evaluation semantics.
The key insight we leverage is that we can instead enrich the demand-driven query evaluation and the incremental edit semantics to dynamically evolve DAIGs such that each step preserves the acyclic dependency structure invariant.
To do so, we use a distinguished edge label (\textsf{fix}) to indicate a dependency on the fixed-point of a given region of the DAIG, which is then dynamically unrolled on-demand by query evaluation and rolled by incremental edits.


The details are formalized in \cref{sec:daig_semantics}; we proceed here by example on the demanded abstract interpretation of the \code{append} program of \cref{fig:overview:cfg}.
%
%
Consider \cref{fig:overview:daig_unroll}, and focus on the black and {\color{black!50!white} grey} vertices and edges, ignoring the {\color{black!40!red}red} for the moment.
Rather than encoding the CFG back edge from $\ell_4$ to $\ell_3$ directly into the DAIG (violating acyclicity in the process), the $\ell_3$-to-$\ell_4$-to-$\ell_3$ loop is initially encoded with separate reference cells for the $0$th and $1$st abstract iterates at the loop head $\ell_3$ (named $\underline{\ell_3}^{\;(0)}$ and $\underline{\ell_3}^{\;(1)}$ respectively), along with a $\textsf{fix}$ edge from those two cells to $\ell_3$'s fixed-point cell $\underline{\ell_3}$, as seen in the {\color{black!50!white} grey} dotted edge.
Crucially, this initial DAIG is acyclic.

\tightpara{Query Evaluation.}
For query evaluation, we can compute fixed points on demand by ``unrolling'' the \emph{abstract interpretation} of loop bodies in the DAIG one {\em abstract} iteration at a time until a fixed point is reached, preserving the acyclic DAIG invariant at each step.
That is, our key observation is to unroll at the semantic level of the abstract interpretation rather than the syntactic level of the control-flow graph.

From the initial DAIG in \cref{fig:overview:daig_unroll}, when the fixed-point $\underline{\ell_3}$ is demanded, its dependencies --- the $0$th and $1$st abstract iterates --- are computed.
If their values are equal, then a fixed-point has been reached and may be written to $\underline{\ell_3}$.\footnote{\label{wideningfootnote}We describe here the widening strategy of applying $\nabla$ every iteration until a fixed-point is reached for simplicity, but the same general idea applies for other widening strategies
or checking convergence with $\sqsubseteq$ instead of $=$.}
If their values are {\em not} equal, then the abstract interpretation of the loop body --- but not the loop body statements --- is unrolled one step further and the \code{fix} edge slides forward to now depend on the $1$st and $2$nd abstract iterates, as seen in the {\color{black!40!red}red} cells and edges of \cref{fig:overview:daig_unroll}. And crucially, this one-step unrolled DAIG is also acyclic.

From here, the process continues, and termination is guaranteed by leveraging the standard argument of abstract interpretation meta-theory: the sequence of abstract iterates in the cells $\underline{\ell_3}^{\;(0)},\underline{\ell_3}^{\;(1)},\underline{\ell_3}^{\;(2)},\dots$ converges because it is produced by widening a monotonically increasing sequence of abstract states, so this demanded unrolling of \textsf{fix} occurs only finitely --- but unboundedly --- many times.
We see that in essence, the sequence of abstract interpretation iterates $\underline{\ell_3}^{\;(0)},\underline{\ell_3}^{\;(1)},\underline{\ell_3}^{\;(2)},\dots$ are encoded into the DAIG on demand during query evaluation.

%

In classical abstract interpretation, a widen $\nabla$ is a join that enforces convergence during interpretation and thus is only strictly needed if the abstract domain has infinite height.
Our approach can be seen as an application of this widening principle to demanded computation.
For an abstract domain of finite height $k$, it would have been sufficient to encode the unrolling of \textsf{fix} eagerly into an acyclic DAIG by inlining the abstract iteration $k$ times to $k$ iterate cells $\underline{\ell_3}^{\;(0)},\dots,\underline{\ell_3}^{\;(k)}$.
However, many expressive, real-world abstract domains --- including the shape analysis domain of our example and most numerical domains  --- are of infinite height.

\tightpara{Incremental Edits.}
Since the acyclic DAIG invariant is always preserved, invalidating on incremental edits still only requires eager dirtying forwards in the DAIG, with some special semantics for \textsf{fix} edges. When dirtying along a \textsf{fix} edge, the \textsf{fix} edge is rolled back to a non-dirty cell (i.e., the $0$th and $1$st iterate). In \cref{fig:overview:daig_unroll}, if the statement cell $\underline{\ell_4}\cdot\underline{\ell_3}$ is edited, then dirtying will happen along the {\color{black!40!red}red} solid \textsf{fix} edge at which point it will slide back to be the {\color{black!50!white} grey} dotted one.

\tightpara{Interprocedural Demand.}
This demanded unrolling of \textsf{fix} also suggests an approach to interprocedural demanded analysis parameterized by a context-sensitivity policy.
To analyze a call when evaluating a query, we construct a DAIG for the callee procedure on demand, indexed by a context determined opaquely by the context-sensitivity policy.

The ``functional approach'' to interprocedural analysis of \citet{sharir1978two} could also potentially be adapted to our framework
by constructing disjoint DAIGs for each phase and inserting dependencies from phase-$2$ callsites to corresponding phase-$1$ summaries.

These techniques both rely on a static call graph, which can be computed soundly using either abstract interpretation (which may itself be expressed in a DAIG) or type-/constraint-based approaches.

%
%

\section{Preliminary Definitions}\label{sec:preliminaries}
Our technique lifts a program and an abstract interpreter \emph{together} into a demanded abstract interpretation graph~(DAIG), a representation that is amenable for sound incremental and demand-driven program analysis.
By design, this construction is {\em generic} in the underlying programming language and concrete semantics as well as the abstract domain and abstract semantics.
In this section, we fix a generic programming language
and an abstract interpreter interface
that serve as inputs to DAIG construction, both to define syntax and to make explicit our assumptions about their semantic properties.
Selected instantiations of the framework for real-world analysis problems are given in \cref{sec:eval}.



\begin{figure}\centering\small
\[\begin{array}{rcl}
  \text{statements} & s\in\mathit{Stmt}\\
  \text{locations} & \ell \in \mathit{Loc}&\\
  \text{control-flow edges} &e\in\mathit{Edge} & ::= \cfedge{\ell}{s}{\ell'}\\
  \text{programs} & \langle L, E, \ell_0\rangle & :~ \mathcal P(\textit{Loc}) \times \mathcal P(\textit{Edge}) \times \textit{Loc}\\
  \text{concrete states} & \sigma\in\Sigma & \text{(with initial state $\sigma_0$)}\\
  \text{concrete semantics} & \denote\cdot & :~ \mathit{Stmt}\to\Sigma\to\Sigma_\bot \\
  \text{collecting semantics} & \denote\cdot^\ast_{\langle L,E,\ell_0\rangle} & :~ \mathit{Loc}\rightarrow\mathcal P(\Sigma)
\end{array}\]
\caption{A generic programming language of control-flow graphs edge-labelled by an unspecified statement language.}
\label{fig:cfglang}
\end{figure}
Programs under analysis are given as control-flow graphs, edge-labelled by an unspecified statement language and interpreted by a denotational concrete semantics as shown in \cref{fig:cfglang}.
A program $\langle L,E, \ell_0\rangle$ is a 3-tuple composed of a set $L$ of control locations, a set $E$ of directed, statement-labelled control-flow edges between locations, and an initial location $\ell_0$.
We say that a program $\langle L,E,\ell_0\rangle$ is {\em well-formed} when
(1) $\ell_0$ and all locations in $E$ are drawn from $L$, and 
(2) $L$ and $E$ form a reducible control-flow graph.
These conditions ensure that we avoid degenerate edge cases and only consider control flow graphs which correspond to realistic programs~\cite{dragon_book}.

Statements are interpreted by the concrete denotational semantics $\denote\cdot$ as partial functions over concrete program states.
As is standard, we can also lift this statement semantics to a collecting 
$\denote\cdot^\ast_{\langle L,E,\ell_0\rangle}$
of full programs, by computing the transitive closure of the statement semantics over a flow graph.
That is, $\denote\ell^*_{\langle L,E,\ell_0\rangle}$ is the set of all concrete states that can be witnessed at program location $\ell$ in a valid program execution.
We elide the subscript when it is clear from context.
Such a collecting semantics is uncomputable in general, but is an important tool for reasoning about analysis soundness.

Now, 
we define the interface of a generic abstract interpreter over this control-flow graph language.
These definitions are intended simply to fix notation and minimize ambiguity and are as standard as possible.

An abstract interpreter is a $6$-tuple $\langle \Sigma^\sharp, \varphi_0, \denote\cdot^\sharp, \sqsubseteq, \sqcup, \nabla \rangle$ composed of:
\begin{itemize}[topsep=0.25em,leftmargin=1em]
\item[-]An {\em abstract domain} $\Sigma^\sharp$ (elements of which are referred to as {\em abstract states}) which forms a semi-lattice under
  \begin{itemize}
  \item[-] a {\em partial order} $\sqsubseteq~\in\mathcal P(\Sigma^\sharp\times\Sigma^\sharp)$
  with a bottom $\bot \in \Sigma^\sharp$
  \item[-] an {\em upper bound} (a.k.a. {\em join}) $\sqcup~:~\Sigma^\sharp\rightarrow\Sigma^\sharp\rightarrow\Sigma^\sharp$
  \end{itemize}
\item[-]An {\em initial abstract state} $\varphi_0\in\Sigma^\sharp$
\item[-]An {\em abstract semantics}  $\denote\cdot^\sharp:\mathit{Stmt}\rightarrow\Sigma^\sharp\rightarrow\Sigma^\sharp$ that interprets program statements as monotone functions over abstract states.

\item[-] A {\em widening} operator $\nabla ~:~\Sigma^\sharp\rightarrow\Sigma^\sharp\rightarrow\Sigma^\sharp$ that is an upper bound operator (i.e., $(\varphi\sqcup\varphi')\sqsubseteq(\varphi\nabla\varphi')$ for all $\varphi,\varphi'$) and enforces convergence (i.e., for all increasing sequences of abstract states $\varphi_0\sqsubseteq\varphi_1\sqsubseteq\varphi_2\sqsubseteq\cdots$, the sequence $\varphi_0,~\varphi_0\nabla\varphi_1,~(\varphi_0\nabla\varphi_1)\nabla\varphi_2,~\dots$ converges).
\end{itemize}
Furthermore, a concretization function $\gamma:\Sigma^\sharp\rightarrow\mathcal P(\Sigma)$ gives meaning to abstract states.
We say that a concrete state $\sigma$ {\em models} an abstract state $\varphi$ (equivalently, that $\varphi$ {\em abstracts} $\sigma$), written $\sigma\models\varphi$, when $\sigma\in\gamma(\varphi)$.
%

\begin{definition}[Local Abstract Interpreter Soundness]
  An abstract interpreter $\aitemplate$ is locally sound if for all $\sigma,\varphi,s$, if $\sigma\models\varphi$ and $\denote s \sigma \neq \bot$ then $\denote s \sigma \models \denote s ^\sharp \varphi$.
\end{definition}


Local soundness can be extended to a global soundness property: if the abstract semantics are locally sound, then the abstract interpreter computes a sound over-approximation of the possible concrete states at each location.

\begin{proposition}[Global Abstract Interpreter Soundness]
\label{thm:aisoundness}
If $\aitemplate$ is locally sound and $\sigma_0\models\varphi_0$ then it induces an abstract collecting semantics ${\denote\cdot^\sharp}^\ast_{\langle L,E,\ell_0\rangle}:\mathit{Loc}\to\Sigma^\sharp$ such that for all $\sigma\in\denote\ell^\ast_{\langle L,E,\ell_0\rangle}$, $\sigma\models{\denote\ell^\sharp}^\ast_{\langle L,E,\ell_0\rangle}$.
\end{proposition}
%
We elide the abstract collecting semantics
${\denote\cdot^\sharp}^\ast_{\langle L,E,\ell_0\rangle}$; it is similarly a transitive closure of the abstract statement semantics over a flow graph.
It is a well-known result that global abstract interpreter soundness is implied by local
soundness~\cite{DBLP:conf/popl/CousotC77}
and that such a global fixed-point is computable using the chaotic iteration method with widening \cite{DBLP:conf/ershov/Bourdoncle93}.


\section{Demanded AI Graphs}
\label{sec:daigs}
Recall from \cref{sec:overview} that a demanded abstract interpretation graph (DAIG) is a directed acyclic hypergraph, whose vertices are reference cells containing program syntax or intermediate analysis results, and whose edges reflect analysis dataflow relationships among those cells.
%
%
%
In \cref{fig:daig_syntax}, we show a syntax for DAIGs.
A DAIG $\mathcal{D}=\langle R,C\rangle$ is composed of a set $R\subseteq\mathit{Ref}$ of named reference cells connected by computation edges $C\subseteq\mathit{Comp}$.
A computation $c\colon n\gets f(n_1,\dots,n_k)$ is an edge connecting sources $\{n_1,\dots,n_k\}$ to a singleton destination $\{n\}$, labeled by some analysis function $f$. 

\begin{figure}\centering
\[\begin{array}{rcl}
  \text{functions} & f&::= \denote\cdot^\sharp ~\vert~ \sqcup ~\vert~ \nabla ~\vert~\textsf{fix} \\
  \text{values} & v &::= s ~\vert~\varphi\\
  \text{names} & n \in \mathit{Nm}&::= \underline\ell ~\vert~ \underline f ~\vert~ \underline i ~\vert~ \underline v ~\vert~ n_1\nameprod n_2 ~\vert~ n^{(i)} \\
  \text{types} & \tau &\in \{\mathit{Stmt},~\Sigma^\sharp\}\\
  \text{reference cells} & r \in \mathit{Ref} &::= n[v:\tau] ~\vert~ n[\varepsilon:\tau]\\
  \text{computations} & c\in\mathit{Comp}  & ::= n \gets f(n_1,\dots,n_k)\\
  \text{DAIGs} & \mathcal{D} & : \mathcal{P}(\mathit{Ref}) \times \mathcal{P}(\mathit{Comp})
\end{array}\]
\caption{Demanded Abstract Interpretation Graphs, edge-labelled by analysis functions and connecting named reference cells storing statements and abstract states.}
\label{fig:daig_syntax}
\end{figure}

Names $\underline\ell$, $\underline f$, $\underline v$ and $\underline i$ correspond respectively to locations $\ell$, functions $f$, values $v$ and integers $i$, supporting memoization of those syntactic constructs.
Name products $n_1\nameprod n_2$ support the construction of more complicated names, and $i$-primed names $n^{(i)}$ allow variants of a single name to be distinguished as loops are unrolled: $n^{(i)}$ is the $i$th unrolled copy of the name $n$ in a loop.
All name equalities are decided structurally.


Values include statements $s$ and abstract states $\varphi\in\Sigma^\sharp$.
Reference cells bind names to values or the absence thereof (denoted $\varepsilon$), while computations specify analysis data-flow dependencies between reference cells.

We denote by $\mathcal D[n\mapsto v]$ the DAIG identical to $\mathcal D$ except that the reference cell named $n$ now holds value $v$.
We also denote DAIG reachability by $n\rightsquigarrow_\mathcal{D} n'$ (eliding the subscript when it is clear from context), and define helper functions \texttt{name}, \texttt{srcs}, and \texttt{dest} to project out, respectively, the name $n$ of a reference cell $n[v_\varepsilon:\tau]$ and the source names $\{n_1,\dots,n_k\}$ or destination name $n$ of a computation $n\gets f(n_1,\dots,n_k)$.
Finally, the typing judgment $R\vdash n\gets f(n_1,\dots,n_k)$ holds when $n_1$ through $n_k$ name references in $R$ with the same types as $f$'s inputs and $n$ names a reference in $R$ with the same type as $f$'s output.

\begin{definition}[DAIG Well-formedness]\label{def:daig_wf}
  A DAIG $\mathcal D = \langle R,C\rangle$ is subject to the following well-formedness constraints.
\begin{itemize}[topsep=0.25em,leftmargin=1.5em]
\item[{\em (1)}] References are named uniquely: \\\phantom{.}\hfill$\forall r,r'\in R~.~\texttt{name}(r) = \texttt{name}(r')~\Leftrightarrow~r = r'$
\item[{\em (2)}] Computations have unique destinations:\\\phantom{.}\hfill$\forall c,c'\in C~.~\texttt{dest}(c) = \texttt{dest}(c')~\Leftrightarrow~c = c'$
\item[{\em (3)}] Dependencies are acyclic:\hfill$\not\exists~r\in R~.~\textsf{name}(r)\rightsquigarrow \textsf{name}(r)$
\item[{\em (4)}] Computations are well-typed with respect to references:\\\phantom{.}\hfill$\forall c\in C~.~R\vdash c$
\item[{\em (5)}] Empty references have dependencies:\\\phantom{.}\hfill $\forall n[\varepsilon:\tau]\in R ~.~\exists~c\in C~.~n=\texttt{dest}(c)$
\end{itemize}
\end{definition}
\begin{figure*}[b]
  \centering
  \input{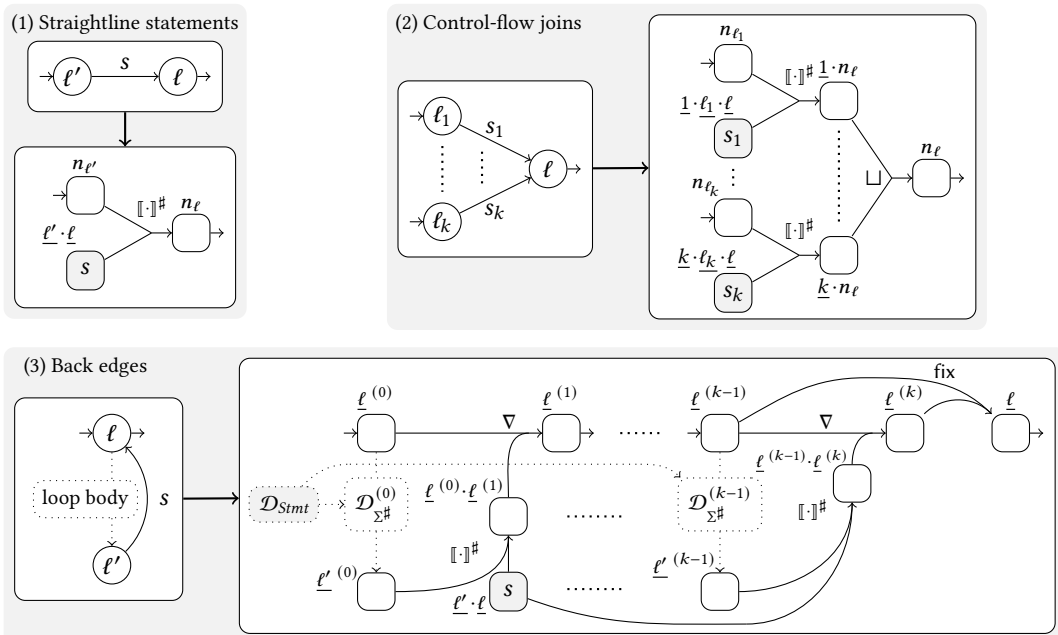}
  \vspace{-0.75em}
  \caption{DAIG--CFG Consistency (\cref{def:cfg_consistency}) in diagram form, illustrating how different CFG structures are encoded into DAIG structures.
    In subfigure (3), we apply some ad-hoc shorthands for the DAIG encoding of the loop body: $\mathcal D_\textit{Stmt}$ contains all of its statement reference cells, while $\mathcal D_{\Sigma^\sharp}^{(i)}$ contains all of its abstract state reference cells, with iteration counts set to $i$.  Each dotted line from $D_\textit{Stmt}$ thus represents one or more DAIG edges, from each statement to corresponding abstract states.
  }
  \label{fig:cfg_daig_snippets}
\end{figure*}

Beyond these basic well-formedness conditions, a DAIG's structure must also properly encode an abstract interpretation computation over an underlying program.
Given a program's CFG and an abstract interpreter interface (as defined in \cref{sec:preliminaries}), there are three general cases shown in \cref{fig:cfg_daig_snippets} to consider when examining a corresponding DAIG.
The key property is that demand-driven query evaluation and incremental edits will evolve the DAIG but preserve the following consistency conditions:

\begin{itemize}[topsep=0.5em,leftmargin=0.5em,itemindent=1em]
\item[(1)]
  A forward CFG edge $\cfedge{\ell'}{s}{\ell}$ to a {\em non-join} location is encoded by a transfer function-labelled DAIG edge, connecting reference cells for its abstract pre-state (named $n_{\ell'}$) and statement label (named $\underline{\ell'}\nameprod\underline\ell$) to a reference cell for its abstract post-state (named $n_\ell$).\footnote{We write $n_\ell$ for the name of the abstract state at $\ell$ throughout this section: $\underline\ell^{\;(0)}$ if $\ell$ belongs to any natural loop and $\underline\ell$ otherwise.  Loop heads $\ell$ are a special case: $n_\ell$ is $\underline\ell^{\;(0)}$ (the abstract state at loop entry) when the destination of a DAIG edge and $\underline\ell$ (the fixed point at $\ell$) otherwise.}
  \item[(2)]
    Forward CFG edges to a {\em join} location $\ell$ are a bit more complex, introducing intermediate cells to encode the join $\sqcup$ into the DAIG.
    For each incoming edge to $\ell$ with statement $s_i$, we introduce a three-cell transfer function construct similar to case (1), with an output cell $\underline i \nameprod n_\ell$ named uniquely for that edge.
    Then, a single join edge connects each pre-join abstract state $\underline i \nameprod n_\ell$ to $n_\ell$, the abstract post-state at $\ell$.
  \item[(3)]
    As described informally in \cref{overview:cycles},
    our framework analyzes CFG back edges by unrolling the abstract fixed-point computation to evolve DAIGs on demand, so this diagram is parameterized by a number $k$ of such unrollings.
    Given the CFG back edge $\cfedge{\ell'}{s}{\ell}$, a transfer function DAIG edge connects the abstract state after one abstract iteration (named $\underline{\ell'}^{\;(0)}$) and $s$ to a pre-widen abstract state at the loop head $\ell$ (named $\underline{\ell}^{\;(0)}\nameprod\underline\ell^{\;(1)}$), which is connected with the previous abstract state at the loop head ($\underline{\ell}^{(0)}$) to the next ($\underline{\ell}^{(1)}$) via a widen edge\footnoteref{wideningfootnote}.
    This acyclic structure is repeated $k$ times in the DAIG (with $k=1$ in the initial construction and further unrollings generated on demand as decribed in \cref{sec:daig_semantics}), thereby encoding the unbounded fixed-point computation.
    Lastly, the \textsf{fix} edge --- indicating a dependency on the eventual fixed point --- connects the two greatest abstract iterates to the reference cell ($\underline \ell$) for the fixed-point abstract state at the loop head.
\end{itemize}
\begin{definition}[DAIG--CFG Consistency]\label{def:cfg_consistency}
  A DAIG $\mathcal D=\langle R,C\rangle$ is consistent with a program CFG $\langle L,E,\ell_0\rangle$, written $\mathcal D \approxeq \langle L,E,\ell_0\rangle$, when it is well-formed and its structure is consistent with that of the program.
\end{definition}
A formal statement of the $\mathcal D\approxeq \langle L,E,\ell_0\rangle$ relation is given in the appendix, closely following the structure of the above description and \cref{fig:cfg_daig_snippets}.

\begin{toappendix}
  First, we fix some notation and terminology related to flow graphs and their structure, all of which is fairly standard:

The edges $E$ of a reducible flow graph program $\langle L,E,\ell_0\rangle$ may be partitioned disjointly into a set $E_f$ of {\em forward} edges and a set $E_b$ of {\em back} edges, where forward edges form a directed acyclic graph that spans $L$ and the destination $\ell'$ of each back edge $\cfedge{\ell}{s}{\ell'}\in E_b$ dominates%
\footnote{A vertex $\ell$ {\em dominates} a vertex $\ell'$ if every path from the entry to $\ell'$ passes through $\ell$.}
the source $\ell$.
A vertex is a {\em loop head} if it is the destination of an edge in $E_b$.

Furthermore, each back edge $e = \cfedge{\ell}{s}{\ell'}\in E_b$ in a reducible flow graph uniquely determines a {\em natural loop} $\textsf{loop}(\ell'):\mathcal P(L)$, the set of locations from which $\ell$ can be reached without passing through $\ell'$.
Intuitively, the natural loop is the ``body'' of the loop with head $\ell'$, and we define $\textsf{loop}(\ell)\triangleq\emptyset$ for all non-loop head locations $\ell$.

There are well-known efficient algorithms both to partition forward/back edges and to find natural loops in reducible flow graphs (e.g., \cite{dragon_book}, Sec. 9.6) which we do not repeat here.

We also define a helper function $\textsf{fwd-edges-to}_{\langle L,E,\ell_0\rangle}:\mathit{Loc}\to\mathcal P(\mathbb N \times \mathit{Edge})$ which 
assigns indices to forward control-flow edges into the given location.
These indices will be essential for DAIG construction, allowing us to uniquely name the intermediate abstract states before a join.

We denote by $L_\sqcup$ the set of CFG join points $\big\{\ell~\big\vert~ \vert\textsf{fwd-edges-to}(\ell)\vert \geq 2\big\}$ and by $L_{\not\sqcup}$ its complement $L/L_\sqcup$. Note that our definition is non-standard: these points are determined by {\em forwards} indegree rather than {\em total} indegree, since no join operation is necessary at a loop entry with only a single non-loop predecessor.

\begin{definition}[DAIG--CFG Consistency]\label{daigcfgconsistencydefn}
  A DAIG $\mathcal D=\langle R,C\rangle$ is consistent with a program CFG $\langle L,E,\ell_0\rangle$, written $\mathcal D \approxeq \langle L,E,\ell_0\rangle$, when it is well-formed and its structure accurately encodes the abstract interpretation of that program by satisfying the following conditions.
  See \cref{fig:cfg_daig_snippets} for a visual representation of this definition to build intuition for the more formal statements below.
  
  \begin{itemize}[topsep=0em,leftmargin=0.5em,itemindent=1em]
  \item[(1)]
    Forward CFG edges to {\em non-join} locations are encoded by a transfer function edge connecting the abstract pre-state and statement to the abstract post-state.
    
    $$\forall \ell'\in L_{\not\sqcup},~\cfedge{\ell}{s}{\ell'}\in E_f\quad .\quad \underline\ell\nameprod \underline{\ell'}[s:\mathit{Stmt}]\in R\quad\wedge\quad n_{\ell'}\gets \denote\cdot^\sharp(n_\ell\nameprod n_{\ell'},n_{\ell})\in C$$
    
  \item[(2)]
    Forward CFG edges to {\em join} locations are slightly more complex: for each join location $\ell$, \textsf{fwd-edges-to} assigns a unique integer index $i$ to each incoming CFG edge.
    For each such CFG edge, a transfer function edge connects the reference cells containing its statement label and the abstract state at its source location to a pre-join abstract state (named by $\underline i \nameprod n_\ell$) specific to that edge.
    Then, a single join edge connects each pre-join abstract state $\underline i \nameprod n_\ell$ to $n_\ell$, the actual abstract state at $\ell$.
    
    $$\forall \ell\in L_\sqcup~.\quad n_\ell\gets\sqcup(\underline 1 \cdot n_\ell,...,\underline{\vert\textsf{fwd-edges-to}(\ell)\vert}\cdot n_\ell)\in C\quad\wedge\quad\left({\begin{array}{c}
        \forall (i,\cfedge{\ell_i}{s_i}{\ell})\in\textsf{fwd-edges-to}(\ell)~.\\
        \underline i\nameprod \underline{\ell_i}\nameprod \underline\ell[s_i:\mathit{Stmt}]\in R~~\wedge~~
        \underline i \nameprod n_\ell\gets \denote\cdot^\sharp(\underline i \nameprod \underline{\ell_i}\nameprod \underline\ell,~n_{\ell'})\in C
    \end{array}}\right)$$
    
  \item[(3)]
    CFG back edges are encoded by some $k$ analysis iterations over the corresponding loop body.
    First, a transfer function edge connects the reference cell (named by $n_{\ell'}=\underline{\ell'}^{\;(0)}$) at the end of the first loop iteration and the back edge's statement label to the first pre-widen abstract state at the loop head (named by $\underline{\ell}^{\;(0)}\nameprod\underline\ell^{\;(1)}$).
    Then, for each $i\in [0, k)$, the loop body is unrolled with iteration count $i$, and widen edges connect the $i$th abstract iterate and pre-widen abstract state to the next abstract iterate.
      Lastly, a \textsf{fix} edge connects the final two abstract iterates to the loop head fixed point $\underline \ell$.

      $$\forall \cfedge{\ell'}{s}{\ell}\in E_b~.~\exists~ k\geq 1~.
      {\begin{array}{l}
      \underline{\ell'}\nameprod\underline\ell[s:\textit{Stmt}]\in R ~~\wedge\\
      \underline\ell^{(0)}\nameprod\underline\ell^{(1)}\gets\! \denote\cdot^\sharp(\underline{\ell'}\!\nameprod\underline\ell, \underline{\ell'}^{(0)})\!\in C~~\wedge\\
      \underline\ell\gets\textsf{fix}(\underline\ell^{(k-1)}\!\!,\underline\ell^{(k)})\in C~~\wedge\\
      \forall i \in [0,k)~.
    \left({\begin{array}{l}
        \underline\ell^{\;(i)}\gets\nabla(\underline\ell^{\;(i-1)},~\underline\ell^{\;(i-1)}\nameprod\underline\ell^{\;(i)})\in C~\wedge\\ 
        \hfill\left\{\textsf{incr-c}^i(c)~\middle\vert~ c\in \textsf{loop-comps}(\ell)\right\}\subseteq C
    \end{array}}\right)\end{array}}$$
  \end{itemize}
  The initial name $n_\ell$ for a location $\ell$ is $\underline\ell^{\;(0)}$ if it belongs to any natural loop and $\underline\ell$ if it does not,
  and we denote by $\textsf{loop-comps}(\ell)$ the subset of $C$ whose elements contain a name $\underline{\ell'}^{\;(0)}$ such that $\ell'\in\textsf{loop}(\ell)$.
  The helper function \textsf{incr} increments the loop-iteration counts in names, and we write \textsf{incr-c} for the pointwise lifting of \textsf{incr} to edges $c\in \mathit{Comp}$.
  \begin{eqnarray*}
\textsf{incr}(n)&\triangleq\left\{{\begin{array}{lr}
    \underline\ell^{\;(k+1)}&\text{ if }n=\underline\ell^{\;(k)}\\
    \textsf{incr}(n_1)\nameprod\textsf{incr}(n_2)&\text{ if }n=n_1\nameprod n_2\\
    n&\text{ otherwise}
\end{array}}\right.
  \end{eqnarray*}
\end{definition}
\end{toappendix}

The above establishes when the structure of a DAIG is consistent with the program's CFG. A DAIG is consistent with an abstract interpret\emph{ation} of the program when the partial analysis results stored in the DAIG are consistent with the partial abstract interpretation, or formally as follows:
\begin{definition}[DAIG--AI Consistency]\label{def:ai_consistency} A DAIG $\mathcal D=\langle R,C\rangle$ is consistent with an abstract interpreter $\aitemplate$, written $\mathcal D \approxeq \aitemplate$, when all partial analysis results stored in $R$ are consistent with the computations encoded by $C$, such that $\underline{\ell_0}[\varphi_0:\Sigma^\sharp]\in R$ and

  \noindent
  $\forall~n[v:\Sigma^\sharp]\in R~,~~n\!\gets\!f(n_1,...,n_k)\in C ~~.\\
    \left\{n_i[v_i\!:\!\tau_i]\middle\vert 1\leq i \leq k\right\}\!\subseteq\! R ~\wedge~ \left\{\!{\begin{array}{lr}v=v_1=v_2&\text{if }f=\textsf{fix}\\v = f(v_1,\dots,v_k)&\!\!\text{otherwise}\end{array}}\right.$
 \end{definition}

\begin{toappendix}
\begin{definition}[Initial DAIG Construction] \label{def:initial_daig} The initial DAIG $\mathcal D_{\small\textsf{init}}(\langle L,E,\ell_0\rangle,\aitemplate)$ for program $\langle L,E,\ell_0\rangle$ and abstract domain $\aitemplate$.
  consists of several distinct categories of both reference cells and computation edges, which we describe and define separately for clarity and ease of presentation.

  Reference cells in $R_\mathit{Stmt}$ contain program statements, those in $R_{\Sigma^\sharp}$ contain abstract states at each control location and join point, and those in $R_\circlearrowleft$ contain those needed to encode loops, while
  computation edges in each $C_f$ encode the uses of that function $f$.
As in \cref{def:cfg_consistency}, we denote by $n_\ell$ the initial name at $\ell$, which is defined as $\underline\ell^{\;(0)}$ if $\ell$ is in the natural loop of some back edge in $E_b$, or $\underline\ell$ otherwise.
  
  That is, $\mathcal D_{\small\textsf{init}}\left(\langle L,E,\ell_0\rangle,\aitemplate\right) \triangleq\left\langle R_\mathit{Stmt}\cup R_{\Sigma^\sharp} \cup R_\circlearrowleft,~C_{\denote\cdot^\sharp}\cup C_\sqcup\cup C_\textsf{fix}\cup C_\nabla\right\rangle$, where:
  \begin{itemize}[topsep=0em,leftmargin=0.5em,itemindent=0.5em]
  \item Reference cells in $R_\mathit{Stmt}$ each contain a program statement and are named by the locations before and after that statement, along with a disambiguating integer index when multiple forwards%
    \footnote{A reducible CFG has at most one back edge to each vertex, so no disambiguation is necessary for back edges.}
    control-flow edges share a destination.
    \begin{align*}
    R_\mathit{Stmt}\triangleq&
    \left\{
    \underline \ell \nameprod\underline {\ell'}[s:\mathit{Stmt}]
    ~\middle\vert~
        (\ell'\in L_{\not\sqcup}~\wedge~\cfedge \ell s \ell' \in E)
        \vee\quad \cfedge{\ell}{s}{\ell'}\in E_b
    \right\}
    \\\cup&\left\{
    \underline i \nameprod\underline \ell\nameprod\underline {\ell'}[s:\mathit{Stmt}]
    ~\middle\vert~
        \ell\in L_{\sqcup}~\wedge~
        (i,\cfedge{\ell'\!}{s}{\ell})\in\textsf{fwd-edges-to}(\ell)
    \right\}
    \end{align*}
  \item $R_{\Sigma^\sharp}$ contains one reference cell of abstract state type per program location, with the initial location $\ell_0$'s cell populated by the initial abstract state $\varphi_0$, along with indexed reference cells for the pre-join abstract states at join point locations.
    \begin{align*}
      R_{\Sigma^\sharp}\triangleq&
      \left\{
      \underline{\ell_0}[\varphi_0:\Sigma^\sharp]\right\}
      \cup
      \left\{
      n_\ell[\varepsilon:\Sigma^\sharp]
      ~\middle\vert~
      \ell\in L/\{\ell_0\}
      \right\}
      \cup 
      \left\{
      \underline i \nameprod n_\ell [\varepsilon:\Sigma^\sharp]
      ~\middle\vert~
      \ell\in L_\sqcup~\wedge~ 1 \leq i \leq |\textsf{fwd-edges-to}(\ell)|
      \right\}
    \end{align*}
  \item Reference cells in $R_\circlearrowleft$ encode the zeroth ($\underline\ell^{\;(0)}$) and first ($\underline\ell^{\;(1)}$) abstract iterates and the first pre-widening abstract state ($\underline\ell^{\;(0)}\nameprod~\underline\ell^{\;(1)}$) at each loop head $\ell$. Names for further iterations' intermediate abstract states are dynamically generated as needed by the operational semantics via demanded unrolling.
    \[
    R_\circlearrowleft\triangleq
    \left\{
    {\underline\ell}^{\;(i)}[\varepsilon:\Sigma^\sharp]
    ~\middle\vert~
    \cfedge{\ell'}{s}{\ell}\in E_b~\wedge~i\in\{0,1\}
    \right\}
    \cup
    \left\{
    \underline\ell^{\;(0)}\nameprod~\underline\ell^{\;(1)}[\varepsilon:\Sigma^\sharp]
    ~\middle\vert~
    \cfedge {\ell'} s \ell \in E_b
    \right\}
    \]

  \item Computations in $C_{\denote\cdot^\sharp}$ encode abstract transfers. Its three subsets respectively encode the abstract semantics of
    forward edges to non-join locations,
    forward edges to join locations,
    and back edges.

    We define the shorthand $\textsf{src-nm}(\ell,\ell')$ to be $\underline{\ell}$ when $\ell$ is a loop head and $\ell'\not\in\textsf{loop}(\ell)$ and $n_\ell$ otherwise.
    \begin{align*}
      C_{\denote\cdot^\sharp} \triangleq&
    \left\{
    n_\ell\gets\denote\cdot^\sharp\left(\underline{\ell'}\nameprod\underline\ell,\textsf{src-nm}(\ell',\ell)\right)
    ~\middle\vert~
    \ell\in L_{\not\sqcup}~\wedge~ \cfedge{\ell'\!}{s}{\ell}\in E_f
    \right\}
    \\
    \cup & \left\{
    \underline i \nameprod n_\ell\gets \denote\cdot^\sharp\left(\underline i \nameprod\underline {\ell'}\nameprod\underline\ell,~\textsf{src-nm}(\ell',\ell)\right)
    ~\middle\vert~
    \ell\in L_\sqcup~\wedge~(i,~\cfedge{\ell'\!}{s}{\ell})\in\textsf{fwd-edges-to}(\ell)
    \right\}
    \\
    \cup & \left\{
    \underline\ell^{\;(0)}\nameprod~\underline\ell^{\;(1)}\gets \denote\cdot^\sharp(\underline{\ell'}\nameprod\underline\ell, n_{\ell'})
    ~\middle\vert~
    \cfedge {\ell'} s \ell \in E_b
    \right\}
    \end{align*}
  \item Computations in $C_\sqcup$ encode joins at program locations of forwards indegree $\geq 2$; the abstract state at $\ell$ is the join of the abstract states on each incoming edge to $\ell$.
    \[
    C_\sqcup\triangleq\left\{
    n_\ell\gets\sqcup(\underline1\nameprod n_\ell,\dots,\underline{k}\nameprod n_\ell )
    ~\middle\vert~
    \ell\in L_\sqcup~\wedge~k=|\textsf{fwd-edges-to}(\ell)|
    \right\}
    \]

  \item Computations in $C_\textsf{fix}$ connect the $0$th and $1$st abstract iterates at a loop head to its fixed point.
    The symbol \textsf{fix} is not a function {\em per se} but rather an indicator of a CFG back edge that is abstractly unrolled by the operational semantics without introducing cyclic dependencies in the static DAIG structure itself.
    \[
    C_\textsf{fix}\triangleq
    \left\{
    \underline\ell \gets \textsf{fix}\left(\underline\ell^{\;(0)},\underline\ell^{\;(1)}\right)
    ~\middle\vert~
    \cfedge{\ell'}{s}{\ell}\in E_b
    \right\}
    \]
  \item Finally, $C_\nabla$ contains widening computations at loop heads; in its initial state, the DAIG includes only the first widen for each loop.
    \[
    C_\nabla\triangleq
    \left\{
    \underline\ell^{\;(1)}\gets\nabla(\underline\ell^{\;(0)},~~\underline\ell^{\;(0)}\nameprod~\underline\ell^{\;(1)})
    ~\middle\vert~
    \cfedge{\ell'} s \ell \in E_b
    \right\}
    \]
  \end{itemize}

\end{definition}
\end{toappendix}

Given a program's CFG and a generic abstract interpretation interface, we can construct an initial DAIG that is consistent with a classical (batch) abstract interpretation: 

\begin{lemma}[Initial DAIG Construction, Well-Formedness, CFG-Consistency, and AI-Consistency] There exists a constructive procedure $\mathcal D_{\small \textsf{\emph{init}}}$ such that for all well-formed programs $\langle L,E,\ell_0\rangle$, the initial DAIG $$\mathcal D = \mathcal D_{\small\textsf{\emph{init}}}(\langle L,E,\ell_0\rangle,\aitemplate)$$ is well formed and consistent with both the target program (i.e. $\mathcal D \approxeq \langle L,E,\ell_0\rangle$) and underlying abstract interpreter (i.e. $\mathcal D \approxeq \aitemplate$). 
\end{lemma}
\begin{proofsketch}
  We define such a $\mathcal D_{\small\textsf{init}}$ in the appendix and show that it produces well-formed results consistent with both the target program and the underlying abstract interpreter.
  Its definition tracks closely with the informal and diagrammatic descriptions above, constructing DAIG structures that correspond to the input CFG.
\end{proofsketch}

\begin{toappendix}
\begin{lemmarep}[Initial DAIG Well-Formedness, CFG-Consistency, and AI-Consistency] For all well-formed programs $\langle L,E,\ell_0\rangle$, $\mathcal D_{\small\textsf{init}}(\langle L,E,\ell_0\rangle)$ is well-formed, $\mathcal D_{\small\textsf{init}}(\langle L,E,\ell_0\rangle)\approxeq\langle L,E,\ell_0\rangle$, and $\mathcal D_{\small\textsf{init}}(\langle L,E,\ell_0\rangle)\approxeq\aitemplate$.
\end{lemmarep}
  \begin{proof}
  DAIG well-formedness holds by construction:
  \begin{itemize}
  \item[(1)] Each component of $R$ constructs reference cell names of a different structure, so no two reference names are equal.
  \item[(2)] Each component of $C$  destination names of a different structure, with the possible exception of $C_\sqcup$, $C_\textsf{fix}$, and the first component of $C_{\denote\cdot^\sharp}$, all of which may construct destination names of the form $\ell$ or $\ell^{\;(0)}$.
    However, since $L_\sqcup$ and $L_{\not\sqcup}$ are disjoint, there is no overlap between the destination names of $C_\sqcup$ and $C_{\denote\cdot^\sharp}$.
    Furthermore, there is no overlap between $C_\textsf{fix}$ and either $C_\sqcup$ or $C_{\denote\cdot^\sharp}$ because if $\cfedge{\ell'}{s}{\ell}\in E_b$ then $n_\ell=\ell^{\;(0)}\neq  \underline\ell$.
  \item[(3)] By construction, $n_\ell\rightsquigarrow n_{\ell'}$ only if there exists a path in $E_f$ from $\ell$ to $\ell'$.  CFG back edges $\cfedge{\ell'}{s}{\ell}$ in $E_b$ maintain this antisymmetry; $n_{\ell'}\rightsquigarrow\underline\ell^{\;(0)}\nameprod\underline\ell^{\;(1)}\rightsquigarrow\underline\ell^{\;(1)}\rightsquigarrow\underline\ell$ through the computations of $C_{\denote\cdot^\sharp}$, $C_\nabla$, and $C_\textsf{fix}$ respectively, and $\underline\ell\not\rightsquigarrow n_{\ell'}$ since $\ell'$ is in the natural loop of $\ell$ and $\underline\ell$ may only appear as the source of edges {\em not} into $\ell$'s natural loop.
  \item[(4)] Each edge in $C$ is well-typed by construction; names of the form $\underline\ell\nameprod\underline{\ell'}$ and and $\underline i \nameprod \underline\ell\nameprod\underline{\ell'}$ are of type $\mathit{Stmt}$ and all others are of type $\Sigma^\sharp$.
  \item[(5)] The name of each empty reference in $R$ appears as the destination of an edge in $C$.
  \end{itemize}
  We elide the details of CFG-consistency, but it is straightforward to verify by cross-referencing the conditions of \cref{def:cfg_consistency} with the set constructions of \cref{def:initial_daig}.  AI-consistency holds vacuously, as the only non-empty reference of type $\Sigma^\sharp$ in $\mathcal{D}_{\small\textsf{init}}$ is $\underline{\ell_0}$, which contains $\varphi_0$.
  \end{proof}
\end{toappendix}

\section{Demanded AI by Evaluating DAIGs}
\label{sec:daig_semantics}
In this section, we give an operational semantics for demand-driven and incremental evaluation of DAIGs.
A state in the operational semantics consists of a DAIG $\mathcal D$ and also an auxiliary memoization table $M$, which can be used to reuse previously-computed analysis results independent of program location.
Memoization tables are finite maps from names $n$ to abstract states $\varphi\in\Sigma^\sharp$, and 
we write $M(n)$ for the abstract state mapped to by $n$ in $M$, \textsf{dom}$(M)$ for the set of names in the domain of $M$, and $M[n\mapsto\varphi]$ for the extension of $M$ with a new mapping from $n$ to $\varphi$.

The DAIG operational semantics are split into two judgments, corresponding to {\em queries} and {\em edits} over both analysis results and program syntax.
Both interaction modes are given in a small-step style, describing the effects of each operation on the DAIG and auxiliary memo table.

\begin{figure}[tb]
\hfill\boxed{\daigevaltemplate}
\vspace{-0.5em}\begin{mathpar}\small
  \inferrule[Q-Reuse]{
    n[v:\tau]\in R
  }{
    \daigeval{\langle R,C\rangle}{M}{n}{v}{\langle R,C\rangle}{M}
  }

  \inferrule[Q-Match]{
    \mathcal D_0=\langle R,C\rangle\\
    n[\varepsilon:\tau]\in R \\
    n\gets f(n_1,\dots,n_k)\in C\\\\
    \daigeval{\mathcal D_{i-1}}{M_{i-1}}{n_i}{v_i}{\mathcal D_i}{M_i} \quad(\text{for }i\in [1, k])\\\\
    \fnname\in\textsf{dom}(M_k)\\
    v=M_k(\fnname)
  }{
    \daigeval{\mathcal D_0}{M_0}{n}{M_k(\fnname)}{\mathcal D_k[n\mapsto M_k(\fnname)]}{M_k}
  }

  \inferrule[Q-Miss]{
    \mathcal D_0=\langle R,C\rangle\\
    n[\varepsilon:\tau]\in R \\
    n\gets f(n_1,\dots,n_k)\in C\\\\
    \daigeval{\mathcal D_{i-1}}{M_{i-1}}{n_i}{v_i}{\mathcal D_i}{M_i} \quad(\text{for }i\in [1,k])\\\\
    \fnname\not\in\textsf{dom}(M_k)\\
    v= f(v_1,\dots,v_k)\\
    f\neq \textsf{fix}
  }{
    \daigeval{\mathcal D_0}{M_0}{n}{v}{\mathcal D_k[n\mapsto v]}{M_k[\fnname \mapsto v]}
  }

  \inferrule[Q-Loop-Converge]{
    n[\varepsilon:\tau]\in R \\
    n\gets \textsf{fix}(n_1,n_2)\in C\\\\
    \daigeval{\langle R,C\rangle}{M}{n_1}{v}{\mathcal D'}{M'}\qquad
    \daigeval{\mathcal D'}{M'}{n_2}{v}{\mathcal D''}{M''}\\
  }{
    \daigeval{\langle R,C\rangle}{M}{n}{v}{D''[n\mapsto v]}{M''}
  }

  \inferrule[Q-Loop-Unroll]{
    n[\varepsilon:\tau]\in R \\
    c=n\gets \textsf{fix}(\underline\ell^{\;(k-1)}, \underline\ell^{\;(k)})\in C\\\\
    \daigeval{\langle R,C\rangle}{M}{\underline\ell^{\;(k-1)}}{v'}{\mathcal D'}{M'}\\
    \daigeval{\mathcal D'}{M'}{\underline\ell^{\;(k)}}{v''}{\mathcal D''}{M''}\\\\
    v'\neq v''\\
    \daigeval{\textsf{unroll}(\mathcal D'',c)}{M''}{n}{v}{\mathcal D'''}{M'''}
  }{
    \daigeval{\langle R,C\rangle}{M}{n}{v}{\mathcal D'''}{M'''}
}
\end{mathpar}
  \caption{
    Operational semantics rules governing {\em queries} for the contents of a DAIG.
    The judgment form $\daigevaltemplate$ is read as ``Requesting $n$ from DAIG $\mathcal{D}$ with auxiliary memo table $M$ yields value $v$, updated DAIG $\mathcal D'$, and updated memo table $M'$.''}
  \label{fig:query_semantics}
\end{figure}

\subsection{Query Evaluation Semantics}
A query for the value of the reference cell with name $n$, given some initial DAIG $\mathcal D$ and auxiliary memo table $M$, yields a value $v$ and (possibly unchanged) DAIG and memo-table structures $\mathcal D'$ and $M'$.
This operation is defined inductively by the $\daigevaltemplate$ judgment form, whose inference rules are given in \cref{fig:query_semantics}.

There are two potential ways to reuse previously-com\-pu\-ted analysis results: either in DAIG $\mathcal D$ or the auxiliary memo table $M$.
The \textsc{Q-Reuse} rule handles the case where the DAIG cell named by $n$ already holds a value, returning that value and leaving the DAIG and memo table unchanged.

The \textsc{Q-Match} and \textsc{Q-Miss} rules handle the case where $n$ is empty in $\mathcal D$.  
In both cases, queries are issued for the input cells $n_1$ through $n_k$ to the computation $f$ that outputs to $n$.
In \textsc{Q-Match}, the auxiliary memo table is matched: $f$ has already been computed for the relevant inputs, so the result is retrieved from $M_k$, the memo table after querying the input cells, and stored in $n$ (i.e., via $\mathcal D_k[n\mapsto M_k(\fnname)]$). Note that this notation is a low-level mutation of the reference cell named by $n$, not an external edit that would trigger invalidation (which we will describe below in \cref{sec:edit_semantics}).

$\textsc{Q-Miss}$ handles memo table misses by computing and memoizing $f(v_1,\dots,v_k)$ before storing the result in both the DAIG $\mathcal D$ and the auxiliary memo table $M$ (i.e., at names $n$ and $\fnname$, respectively).


\subsection{Demanded Fixed Points}\label{sec:unrolling}
Demanded unrolling is the process by which we compute abstract interpretation fixed-points over cyclic control flow graphs without introducing cyclic dependencies into DAIGs, as described informally in \cref{sec:overview} and represented graphically in Fig.~\ref{fig:overview:daig_unroll}.
The semantics are formalized by the \textsc{Q-Loop-Converge} and \textsc{Q-Loop-Unroll} rules in \cref{fig:query_semantics}.

Recall that \textsf{fix} is a special function symbol indicating an analysis fixed-point computation.
The destination of a \textsf{fix} edge is a loop-head cell for storing a fixed-point invariant, and its sources are the two greatest abstract iterates of said loop head yet computed.
When those abstract iterates have the same value $v$, the analysis has reached a fixed point\footnoteref{wideningfootnote}, and a query for the fixed point may return $v$.
However, when they are unequal, a query triggers an unrolling in the DAIG: the loop body's abstract state reference cells are unrolled one more iteration, the \textsf{fix} edge's sources are shifted forward one iteration, and then the fixed-point query is reissued. 

This procedure differs from concrete/syntactic loop unrolling --- e.g. as applied by an optimizing compiler or bounded model checker --- in that it applies to the DAIG's reified abstract interpretation computation, including joins and widens, \emph{not} to the concrete syntax of the program under analysis.
That is, the $k$-th demanded unrolling corresponds to the $k$-th application of the loop body's abstract semantics (i.e. the $k$-th abstract iteration) rather than the $k$-th concrete execution of the loop.
As a result, it is sound with respect to the concrete semantics of the program under analysis and is guaranteed
to converge.

As the name suggests, \textsc{Q-Loop-Converge} applies when the abstract interpretation has reached a fixed point.
Since the dependencies of the \textsf{fix} edge, $n_1$ and $n_2$, are consecutive abstract iterates at the head of the corresponding loop, their evaluation to the same value $v$ indicates that loop analysis has converged, so $v$ may be stored in the DAIG and returned.
%

On the other hand, \textsc{Q-Loop-Unroll} applies when the abstract interpretation has not yet reached a fixed point, since the two most recent abstract iterates are unequal.
In this case, the loop is unrolled once by the \textsf{unroll} helper function and the query for the fixed point is reissued.
The \textsf{unroll} helper function used in \textsc{Q-Loop-Unroll} takes a DAIG $\mathcal D$ and a \textsf{fix} edge and unrolls the loop corresponding to the \textsf{fix} edge by one iteration in $\mathcal D$.
It is defined as follows:
\[\begin{tabular}{@{}l@{}}
$\textsf{unroll}\left(\langle R,C\rangle, c = \underline\ell\gets \textsf{fix}(\underline\ell^{\;(k-1)}, \underline\ell^{\;(k)})\right) \triangleq \langle R',C'\rangle\text{, where}$
\\
\quad$R'=R\cup\left\{\textsf{incr}(n)[\varepsilon:\Sigma^\sharp] ~\middle\vert~ \underline\ell^{\;(k-1)}\!\rightsquigarrow n\rightsquigarrow\! \underline\ell^{\;(k)} \right\}$ and
\\
\quad$C'= C/ \{c\}\cup\left\{\underline\ell\gets \textsf{fix}\left(\underline\ell^{\;(k)}, \underline\ell^{\;(k+1)}\right)\right\}$
\\
\quad$\phantom{C'=C/ \{c\}}\cup\left\{\textsf{incr-c}(c)~\middle\vert~ \underline\ell^{\;(k-1)}\!\rightsquigarrow\!\texttt{dest}(c)\!\rightsquigarrow\! \underline\ell^{\;(k)} \right\}$
\end{tabular}\]
where \textsf{incr} and \textsf{incr-c} increment the iteration counts of names.
Intuitively, \textsf{unroll} takes the region of the DAIG forwards-reachable from the $k-1^{th}$ abstract iterate $\underline\ell^{\;(k-1)}$ and backwards-reachable from the $k^{th}$ abstract iterate $\underline\ell^{\;(k)}$ and duplicates it while incrementing all name's iterations counts from $k\!-\!1$ to $k$, then shifts the \textsf{fix} edge forward one iteration.
Crucially, this operation preserves the DAIG acyclicity invariant.

\subsection{Incremental Edit Semantics}
\label{sec:edit_semantics}
An \emph{edit} to a DAIG $\mathcal D$ occurs when a value $v$ is written to some reference cell named $n$ in $\mathcal D$ by an external mutator.  This edit must both update $n$ and also clear the value of (or ``dirty'') any reference cell that (transitively) depends on $n$.  Here we give a full definition of the edit operation, using the $\daigedittemplate$ judgment given in \cref{fig:edit_semantics}.

\begin{figure}[tb]
  \hfill\boxed\daigedittemplate
\vspace{-0.5em}\begin{mathpar}\small
  \inferrule[E-Commit]
            {\forall c\in C ~.~n\in\texttt{srcs}(c)\implies \texttt{dest}(c)[\varepsilon:\tau]\in R\\\\
              \underline\ell\gets\textsf{fix}(\underline\ell^{\;(i)}, \underline\ell^{\;(i+1)})\in C\implies \underline\ell^{\;(1)}[\varepsilon:\tau]\in R\\\\
              v_\varepsilon=\varepsilon ~\implies~\exists c\in C~.~n=\texttt{dest}(c)\\\\
              v_\varepsilon\not=\varepsilon ~\implies~\exists n[\text{\textunderscore}:\tau]\in R ~.~ v_\varepsilon : \tau
            }
            {\daigedit{\langle R,C\rangle}{n}{v_\varepsilon}{\langle R,C\rangle[n\mapsto v_\varepsilon]}}

  \inferrule[E-Propagate]
            {n\rightsquigarrow_{\mathcal D} n'\\
              \daigedit{\mathcal D}{n'}{\varepsilon}{\mathcal D'}\\
              \daigedit{\mathcal D'}{n}{v_\varepsilon}{\mathcal D''}
            }
            {\daigedit{\mathcal D}{n}{v_\varepsilon}{\mathcal D''}}

  \inferrule[E-Loop]
            {\underline\ell\gets \textsf{fix}(\underline\ell^{\;(k-1)}, \underline\ell^{\;(k)})\in C\\
               \daigedit{\langle R,C'\rangle}{\underline\ell^{\;(1)}}{\varepsilon}{\mathcal D'}\\\\
              C' = C / \left\{\underline\ell\gets \textsf{fix}(\underline\ell^{\;(k-1)}, \underline\ell^{\;(k)})\right\}\cup\left\{\underline\ell\gets \textsf{fix}(\underline\ell^{\;(0)}, \underline\ell^{\;(1)})\right\}
            }
            {\daigedit{\langle R,C\rangle}{\underline\ell^{\;(k)}}{\varepsilon}{\mathcal D'}}
\end{mathpar}
  \caption{Operational semantics rules governing {\em edits} to the contents of a DAIG. The judgment $\daigedittemplate$ is read as ``Editing reference cell $n$ of DAIG $\mathcal D$ with value $v_\varepsilon$ yields updated DAIG $\mathcal D'$,'' where $v_\varepsilon$ ranges over values $v$ and the ``empty'' symbol $\varepsilon$.}
  \label{fig:edit_semantics}
\end{figure}

As described informally in \cref{sec:overview}, invalidation proceeds by dirtying forwards in the acyclic DAIG, except that the implicit cyclic dependency from \textsf{fix} edges must be accounted for, by rolling back to a non-dirty source cell.


The \textsc{E-Commit} rule is a base case: if the edited cell's downstream dependencies are all empty, then the edit may be performed directly.
Its second premise accounts for the implicit dependency of abstract iterate cells
$\underline{\ell}^{\;(i)}$ for $i > 0$ on the fixed point cell $\underline{\ell}$ (corresponding to the loop back edge in the control-flow graph); it suffices to check that the $1$st abstract iterate cell has been emptied, as all abstract iterates $\underline{\ell}^{\;(i)}$ for $i > 1$ are reachable from $\underline{\ell}^{\;(1)}$.
The third and fourth premises ensure that DAIG well-formedness is preserved, by preventing emptying of source nodes and ill-typed edits respectively.

The \textsc{E-Propagate} rule recursively empties reference cells that depend on the edited cell, eventually bottoming out when no such cells are non-empty and \textsc{E-Commit} can be used to derive the $\daigedit{\mathcal D'}{n}{v_\varepsilon}{\mathcal D''}$ premise. Note that there is no recomputation here in \textsc{E-Propagate}, only emptying.


The \textsc{E-Loop} rule applies when the final abstract iterate of a loop is dirtied.
The sources of its \textsf{fix} edge are reset to its $0$th and $1$st abstract iterates and dirtying continues from the $1$st abstract iterate.
This handling is slightly more conservative than necessary in the case that an intermediate abstract iterate (i.e., with $k > 1$) is edited since results from some previous iterations (up to that $k$) may not need to be discarded,
but it simplifies the presentation and handles all program edits with maximal reuse.

\section{Soundness, Termination, and From-Scratch Consistency}
\label{sec:metatheory}

In this section, we state two key properties of demanded abstract interpretation graphs:
{\em from-scratch consistency}, which guarantees that DAIG query results are identical to the analysis results computed by the underlying abstract interpreter at a global fixed-point,
and {\em query termination}, which guarantees termination of the DAIG query semantics even in the presence of unbounded abstract loop unrolling.
The proofs are deferred to the appendix due to space constraints.

Both theorems rely on the preservation of DAIG well-formedness (\cref{def:daig_wf}), DAIG--CFG consistency (\cref{def:cfg_consistency}), and DAIG--AI consistency (\cref{def:ai_consistency}) under queries and program edits.
\newcommand{\daigprogramedit}{\daigedit{\mathcal D}{n}{s}{\mathcal D'}}
\begin{lemma}[DAIG Well-Formedness Preservation]\label{thm:wf_preservation}~\\If $\mathcal D$ is well-formed and either $\daigevaltemplate$ or $\daigedittemplate$, then $\mathcal D'$ is well-formed.
\end{lemma}
\begin{toappendix}
\begin{lemmarep}[DAIG Well-Formedness Preservation]\label{thm:wf_preservation} If $\mathcal D$ is well-formed and either $\daigevaltemplate$ or $\daigedittemplate$, then $\mathcal D'$ is well-formed.
\end{lemmarep}
\begin{proof}
  By structural induction on the operational semantics derivation.

  Note that the only DAIG well-formedness condition that refers to reference cells contents is (5), which is trivially preserved by each query rule \textsc{Q-}$\ast$ since they never empty reference cells.
  Edit rule \textsc{E-Commit}'s third premise ensures that edits don't violate (5), while \textsc{E-Loop} and \textsc{E-Propagate} only affect cells with non-zero indegree and therefore also preserve (5).
  Thus, only those rules that affect the {\em structure} of the DAIG must be considered: \textsc{Q-Loop-Unroll} and \textsc{E-Loop}.

  \hangindent=\parindent
  \underline{Case \textsc{Q-Loop-Unroll}:} By induction, $\mathcal D'$ and $\mathcal D''$ are well-formed.
  The \textsf{unroll} procedure preserves well-formedness
  condition (1) since $incr(n)\neq n$ for each $n$ in the set of added reference cells $\left\{\textsf{incr}(n)[\varepsilon:\Sigma^\sharp] ~\middle\vert~ \underline\ell^{\;(k-1)}\!\!\!\rightsquigarrow n\rightsquigarrow\! \underline\ell^{\;(k)} \right\}$ as each such $n$ is an abstract state in a loop body and therefore has an iteration count;
  condition (2) since $c$ is removed from C before it is replaced and by the same argument as (1) for the destination of each edge in $\left\{\textsf{incr-c}(c)~\middle\vert~ \underline\ell^{\;(k-1)}\!\!\!\rightsquigarrow\!\texttt{dest}(c)\!\rightsquigarrow\! \underline\ell^{\;(k)} \right\}$;
  conditions (3) and (4) by isomorphism from the previous iteration's edges to the unrolled iteration's edges;
  and condition (5) since each added reference has a corresponding added incoming edge.

  Therefore, $\textsf{unroll}(\mathcal D'',c)$ is well-formed, so by induction $\mathcal D'''$ is well-formed.

  \underline{Case \textsc{E-Loop}:}
  Except for the \textsf{fix} edge whose inputs are changed, the intermediate DAIG $\langle R,C'\rangle$ is identical to $\mathcal D$, which is well-formed.
  Therefore, $\langle R,C'\rangle$ trivially satisfies well-formedness conditions (1) and (3), which refer only to references.
  It satisfies conditions (2) and (5) since each reference in $R$ has the same number of incoming edges in $C$ and $C'$ because the added and removed edge have the same destination, and condition (4) since the sources of both the added and removed edge are of type $\Sigma^\sharp$.
  Therefore, $\langle R,C'\rangle$ is well-formed, so by induction $\mathcal D'$ is well-formed.
  
\end{proof}
\end{toappendix}

\begin{lemma}[DAIG--CFG Consistency Preservation]\label{thm:cfg_consistency_preservation}~\\If $\mathcal D\approxeq \langle L,E,\ell_0\rangle$ then:
  \begin{itemize}
  \item[-]if $\daigevaltemplate$ then $\mathcal D'\approxeq \langle L,E,\ell_0\rangle$;
  \item[-]if $\daigprogramedit$ then $\mathcal D'\approxeq \langle L,E',\ell_0\rangle$, where $E'$ is $E$ with the edit applied (see \cref{thm:cfg_consistency_preservation_rep} for details).
  \end{itemize}
\end{lemma}
\begin{toappendix}
\begin{lemmarep}[DAIG--CFG Consistency Preservation]\label{thm:cfg_consistency_preservation_rep} If $\mathcal D\approxeq \langle L,E,\ell_0\rangle$ then:
  \begin{itemize}
  \item[-]if $\daigevaltemplate$ then $\mathcal D'\approxeq \langle L,E,\ell_0\rangle$;
  \item[-]if $\daigprogramedit$ then $\mathcal D'\approxeq \langle L,E',\ell_0\rangle$, where $E'$ is $E$ with the edit applied.
  \end{itemize}
\end{lemmarep}
The first bullet point states that DAIG--CFG consistency is preserved under DAIG evaluation, while the second states that a program edit ({\em not} an arbitrary edit to abstract state) in a DAIG consistent with an initial CFG yields a DAIG consistent with the corresponding edited CFG.

The post-edit CFG is defined as follows: since $n$ names a reference cell of type $\mathit{Stmt}$ in $\mathcal D$ and $\mathcal D\approxeq \langle L,E,\ell_0\rangle$, $n$ is either of the form $\underline i \nameprod \underline \ell \nameprod \underline \ell'$ or $\underline \ell \nameprod \underline \ell'$, said reference cell holds some non-$\varepsilon$ value $s':\mathit{Stmt}$, and $\cfedge{\ell}{s'}{\ell'}\in E$.
Thus, the post-edit CFG edges are $E'\triangleq E / \{\cfedge{\ell}{s'}{\ell'}\}\cup\{\cfedge{\ell}{s}{\ell'}\}$.  That is, $E'$ is $E$ with whichever edge corresponds to DAIG name $n$ re-labelled by $s$.

Note that this lemma as stated applies only to in-place edits of CFG edges, but the same technique applies for deletions (which can be seen as in-place edits where $s$ is the no-op \texttt{skip} statement) and insertions (for which a new sub-DAIG is constructed and inserted, before dirtying proceeds as usual).

\begin{proof}[Proof (Bullet 1):] By cases on the derivation $\daigevaltemplate$.
  All cases other than \textsc{Q-Loop-Unroll} are trivial since they don't modify the structure of the CFG.
  
  \textsc{Q-Loop-Unroll} abstractly unrolls a loop body by adding DAIG reference cells and computations with \textsf{unroll}.  Since \textsf{unroll} increments the iteration counts of names in the new unrolling, conditions (1) and (2) of \ref{def:cfg_consistency} are vacuously satisfied as they only reference the initial (i.e. zeroth) copy of loop body reference cells.
  Condition (3) requires that for each \textsf{fix} edge, there exist unrollings of the corresponding loop body's DAIG region for all iterations up to that fix edge's sources.  The $i=0$ through $i=k-1$ cases are satisfied in the pre-unrolling DAIG, and \textsf{unroll} adds the $k$th unrolling by incrementing each reference in the $k-1$th, so \textsc{Q-Loop-Unroll} preserves DAIG--CFG consistency as well.
\end{proof}
\begin{proof}[Proof (Bullet 2):]
  The proof is straightforward: the only possible {\em structural} changes from the initial DAIG $\mathcal D$ to the post-edit DAIG $\mathcal D'$ are \textsf{fix} edges reset to $k=1$ by rule \textsc{E-Loop}, which preserve condition (3) of definition \cref{daigcfgconsistencydefn} for any back edges reachable from the edit.
  The only possible reference cell {\em content} changes from $\mathcal D$ to $\mathcal D'$ are dirtied abstract state cells (which don't affect \cref{daigcfgconsistencydefn}) and the edited statement cell $r=n[s:\textit{Stmt}]$, whose value $s$ agrees with the corresponding CFG edge in $E'$ and thus satisfies the relevant $r\in R$ conjunct of whichever condition of \cref{daigcfgconsistencydefn} applies to its edge type (i.e. straightline forward edge, join-point forward edge, or back edge).
\end{proof}

\end{toappendix}

\begin{lemma}[DAIG--AI Consistency Preservation]\label{thm:ai_consistency_preservation}~\\If $\mathcal D\approxeq \aitemplate$ and either $\daigevaltemplate$ or $\daigprogramedit$, then $\mathcal D'\approxeq \aitemplate$.
\end{lemma}
\begin{toappendix}
\begin{lemmarep}[DAIG--AI Consistency Preservation]\label{thm:ai_consistency_preservation} If $\mathcal D\approxeq \aitemplate$ and either $\daigevaltemplate$ or $\daigprogramedit$, then  $\mathcal D'\approxeq \aitemplate$.
\end{lemmarep}
\begin{proof}
  The two cases in the premise of the lemma correspond to query evaluation and program edits.

  For program edits, DAIG--AI consistency is clearly preserved:
  \begin{itemize}
  \item[-] Reference cell $\underline{\ell_0}$ in $\mathcal D'$ is unchanged from $\mathcal D$ since it is not reachable from any statement-type reference cell (which $n$ is guaranteed to be by the well-typedness premise of rule \textsc{E-Commit}).
  \item[-] All in-degree $\geq 1$ reference cells in $\mathcal D'$ are either empty (in which case they vacuously satisfy Def.~\ref{def:ai_consistency}) or non-empty (in which case they satisfy Def.~\ref{def:ai_consistency} for $\mathcal D'$ because they did so for $\mathcal D$).
    \end{itemize}
  
  For query evaluation, we proceed by cases on the derivation of $\daigevaltemplate$.
  \begin{itemize}
  \item[-] The \textsc{Q-Reuse} and \textsc{Q-Loop-Unroll} cases are trivial because they do not change reference cells contents or add non-empty reference cells.
  \item[-] \textsc{Q-Miss}'s premises are identical to \cref{def:ai_consistency}'s $f\neq\textsf{fix}$ case, and \textsc{Q-Match}'s premises, given that $M_k(\fnname)=f(v_1,\dots,v_k)$ by memoization table soundness, are the same.
  \item[-] \textsc{Q-Loop-Converge}'s premises are identical to \cref{def:ai_consistency}'s $f=\textsf{fix}$ case.
  \end{itemize}
\end{proof}
\end{toappendix}

With these preservation results, we can now prove that DAIG query results for abstract states at program locations are from-scratch consistent with the global fixed-point invariant map of the DAIG's underlying abstract interpreter.

\begin{theorem}[DAIG From-Scratch Consistency]\label{thm:fsc}
  For all sound $M$ and well-formed $\mathcal D$ such that $\mathcal D\approxeq\langle L,E,\ell_0\rangle$ and $\mathcal D\approxeq\aitemplate$, if $\daigeval{\mathcal D}{M}{\underline\ell}{v}{\mathcal D'}{M'}$ then $v = {\denote\ell^\sharp}^\ast_{\langle L,E,\ell_0\rangle}$.
\end{theorem}
\begin{toappendix}
\begin{theoremrep}[DAIG From-Scratch Consistency]\label{thm:fsc}
  For all sound $M$ and well-formed $\mathcal D$ such that $\mathcal D\approxeq\langle L,E,\ell_0\rangle$ and $\mathcal D\approxeq\aitemplate$, if $\daigeval{\mathcal D}{M}{\underline\ell}{v}{\mathcal D'}{M'}$ then $v = {\denote\ell^\sharp}^\ast_{\langle L,E,\ell_0\rangle}$.
\end{theoremrep}
\begin{proof}
  If the reference cell named by $\underline\ell$ is non-empty in $\mathcal D$, then \textsc{Q-Reuse} must be the root of the derivation $\daigeval{\mathcal D}{M}{\underline\ell}{v}{\mathcal D'}{M'}$.  Therefore, since $\mathcal D\approxeq\aitemplate$, $v=\denote{\ell}^\sharp{}^\ast$.

  Otherwise, the reference cell named by $\underline\ell$ is empty in $\mathcal D$ and we will proceed
  by cases on $\ell$, a non-join location $\ell\in L_{\not\sqcup}$ with CFG in-degree 1, a join location $\ell\in L_\sqcup$ with CFG in-degree $\geq 2$, or a loop head $\ell$ such that there exists a back edge $\cfedge{\ell'}{s}{\ell}\in E_b$.  Note that $\ell_0$ --- the only location that doesn't fall into one of those cases --- is excluded because is the reference cell with name $\underline{\ell_0}$ is non-empty by $\mathcal D \approxeq \aitemplate$.
  
  \hangindent=\parindent
  \underline{{\em Case} $\ell\in L_{\not\sqcup}$:}\hspace{0.5em} Since $\mathcal{D}\approxeq\langle L,E,\ell_0\rangle$, there exists an $\cfedge{\ell'}{s}{\ell'}\in E_f$, $\underline{\ell'}\nameprod\underline{\ell}[s:\textit{Stmt}]\in R$, and $\underline\ell\gets \denote\cdot^\sharp(\underline{\ell'}\nameprod\underline\ell,\underline{\ell'})\in C$.
  Therefore, the root of the derivation of $\daigeval{\mathcal D}{M}{\underline\ell}{v}{\mathcal D'}{M'}$ is either \textsc{Q-Miss} or \textsc{Q-Match}.
  
  \parindent=2em
  \hangindent=\parindent
  \underline{{\em Sub-case} \textsc{Q-Miss}:}
  $n_1=\underline{\ell'}\nameprod\underline\ell$ and therefore $v_1=s$ by \textsc{Q-Reuse}.
  Due to the preceding preservation lemmas and the inductive hypothesis, $\daigeval{\mathcal D_1}{M_1}{\underline{\ell'}}{v_2}{\mathcal D_2}{M_2}$ and $v_2=\denote{\ell'}^\sharp{}^\ast$.
  Therefore, $v= \denote s^\sharp\denote{\ell'}^\sharp{}^\ast$ which, by local abstract interpreter soundness, is equal to $\denote{\ell}^\sharp{}^\ast$.

  \parindent=2em
  \hangindent=\parindent
  \underline{{\em Sub-case} \textsc{Q-Match}:}
  By the same arguments as for the \textsc{Q-Miss} case --- which follow from premises shared between the two rules --- $v_1=s$ and $v_2=\denote{\ell'}^\sharp{}^\ast$.
  Therefore, by memoization table soundness, $v=M_2(\underline {\denote\cdot^\sharp}\nameprod \underline s\nameprod \underline {\denote{\ell'}^\sharp{}^\ast})= \denote s ^\sharp \denote{\ell'}^\sharp{}^\ast$ which, by local abstract interpreter soundness, is equal to $\denote\ell^\sharp{}^\ast$.
  
  \parindent=1em
  \hangindent=\parindent
  \underline{{\em Case} $\ell\in L_{\sqcup}$:}\hspace{0.5em} Since $\mathcal{D}\approxeq\langle L,E,\ell_0\rangle$, there are $k$ forward CFG edges into $\ell$ and for each such $\cfedge {\ell_i}{s_i}{\ell}$, we have $\underline i \cdot \underline \ell_i \cdot\underline \ell[s_i:\textit{Stmt}]\in R$ and $\underline i \cdot \underline \ell\gets \denote \cdot ^\sharp (\underline i \nameprod \underline \ell_i\nameprod\underline \ell,~\underline \ell_i)\in C$.
  Also by $\mathcal{D}\approxeq \langle L,E,\ell_0\rangle$, there is an edge $\underline\ell \gets \sqcup(\underline 1\nameprod \underline \ell, \dots, \underline k \nameprod \underline\ell)\in C$.
  Therefore, the root of the derivation of $\daigeval{\mathcal D}{M}{\underline\ell}{v}{\mathcal D'}{M'}$ is either \textsc{Q-Miss} or \textsc{Q-Match}.
  We omit the \textsc{Q-Match} subcase as it is analogous to the \textsc{Q-Miss} subcase in the same manner as the previous ($\ell\in L_{\not\sqcup}$) case.

  \parindent=2em
  \hangindent=\parindent
  \underline{{\em Sub-case} \textsc{Q-Miss}:}
  For each $i$, $n_i$ = $\underline i \nameprod \underline \ell$.
  Since $\underline i\nameprod \underline\ell_i\nameprod\underline\ell[s_i:Stmt]\in R$ and $\underline i \nameprod \underline \ell \gets \denote\cdot^\sharp (\underline i \nameprod \underline \ell_i\nameprod\underline\ell,~\underline\ell_i)\in C$, we know that $v_i=\denote {s_i}^\sharp\denote{\ell_i}^\sharp{}^\ast$, as the subquery for $\underline i \nameprod \underline \ell_i\nameprod\underline\ell$ resolves to $s_i$ via \textsc{Q-Reuse} and the subquery for $\underline\ell_i$ resolves to $\denote{\ell_i}^\sharp{}^\ast$ by the inductive hypothesis.
  Then, $v=\sqcup(\denote{s_1}^\sharp\denote{\ell_1}^\sharp{}^\ast,\dots,\denote{s_k}^\sharp\denote{\ell_k}^\sharp{}^\ast)$ which by global abstract interpreter soundness is equal to $\denote\ell^\sharp{}^\ast$ --- the invariant at $\ell$ is the join of each of its predecessor's invariants, transformed by the abstract semantics of the CFG edge by which they're connected.

  \parindent=1em
  \hangindent=\parindent
  \underline{{\em Case} $\ell$ is a loop head:}\hspace{0.5em}
  Since $\mathcal D\approxeq\langle L,E,\ell_0\rangle$, $\ell$ is the destination of a CFG back-edge $\cfedge{\ell'}{s}{\ell}\in E_b$ and $\underline\ell$ is the destination of a DAIG \textsf{fix} edge $\underline\ell \gets \textsf{fix}(\underline\ell^{\;(k-1)},\underline\ell^{\;(k)})\in C$.
  
  By natural number induction, a query for $\underline\ell^{\;(i)}$ yields the  $i$th abstract iterate at $\underline\ell$.
  In the $i=0$ base case, a query for $\underline\ell^{\;(0)}$ yields the $0$th abstract iterate at $\ell$ by precisely the argument%
  \footnote{Modulo the change of name; note that the initial name $n_\ell$ for a loop head $\ell$ is $\underline\ell^{\;(0)}$ as opposed to $\underline\ell$.}
  of the previous two cases, which deal with analysis of forwards dataflow.
  In the inductive case, 
  a query for $\underline\ell^{\;(k)}$ yields the $k$th abstract iterate at $\ell$, since $\underline\ell^{\;(k)}\gets \nabla(\underline\ell^{\;(k-1)}, \underline\ell^{\;(k-1)}\nameprod\underline\ell^{\;(k)})\in C$ by $\mathcal D\approxeq\langle L,E,\ell_0\rangle$, and a query for $\underline\ell^{\;(k-1)}\nameprod\underline\ell^{\;(k)}$ yields the image of the $k$th abstract iterate in the abstract semantics of the loop body --- as unrolled by definition of $\textsf{unroll}$ --- by local abstract interpreter soundness and $\mathcal D\approxeq\aitemplate$.

  The root of the derivation tree of $\daigeval{\mathcal D}{M}{\underline\ell}{v}{\mathcal D'}{M'}$ is either \textsc{Q-Loop-Converge}
  --- in which case the result $v$ of the subqueries for the two most recent abstract iterates are equal and $v$ is therefore the fixed-point $\denote\ell^\sharp{}^\ast$ ---
  or \textsc{Q-Loop-Unroll}, in which case $v$ is equal to $\denote\ell^\sharp{}^\ast$ by the inductive hypothesis.
\end{proof}
\end{toappendix}

\begin{corollary}
  Query results are sound.
\end{corollary}

Since the global invariant map $\denote\cdot^\sharp{}^\ast$ of the underlying abstract interpreter $\aitemplate$ is sound (by Global Abstract Interpreter Soundness (\cref{thm:aisoundness}))
and a DAIG query for the abstract state at a location $\ell$ returns $\denote\ell^\sharp{}^\ast$, DAIG query results themselves are sound.

\begin{theorem}[DAIG Query Termination]\label{thm:termination} For all $M$ and well formed $\mathcal D$ such that $\mathcal D\approxeq\langle L,E,\ell_0\rangle$ and $\mathcal D\approxeq\aitemplate$, if $n$ is in the namespace of $\mathcal D$ then there exist $v,\mathcal D', M'$ such that $\daigevaltemplate$.
\end{theorem}

\begin{toappendix}
\begin{theoremrep}[DAIG Query Termination]\label{thm:termination} For all $M$ and well formed $\mathcal D$ such that $\mathcal D\approxeq\langle L,E,\ell_0\rangle$ and $\mathcal D\approxeq\aitemplate$, if $n$ is in the namespace of $\mathcal D$ then there exist $v,\mathcal D', M'$ such that $\daigevaltemplate$.
\end{theoremrep}
\begin{proof}
  Let the ancestors of a name $n$ in $\mathcal D$ be the set of backwards-reachable names  $\{n'~\vert~n'\rightsquigarrow n\}$.
  Note that, since $\mathcal D$ is well formed and therefore acyclic, if $n\gets f(n_1,\dots,n_k)\in C$ then each $n_i$ has strictly fewer ancestors than $n$.
  We will proceed by induction on the number of ancestors of $n$.

  \hangindent=\parindent
  \underline{Base case:}\hspace{0.5em}
  Since $n$ has no ancestors, there is no $c\in C$ such that $\texttt{dest}(c)=n$.
  Thus, by definition 4.1.5, $n$ names a non-empty reference cell (i.e. $n[v:\tau]\in R$).
  Therefore, by \textsc{Q-Reuse},  $\daigeval{\mathcal D}{M}{n}{v}{\mathcal D}{M}$.

  \hangindent=\parindent
  \underline{Inductive case:}\hspace{0.5em}
  By well-formedness of $\mathcal D$ (definition 4.1.2), there is exactly one edge $c=n\gets f(n_1,\dots, n_k)\in C$ with destination $n$.
  We proceed by cases on $f$.

  \parindent=2em
  \hangindent=\parindent
  \underline{{\em Case} $f\in\{\denote\cdot^\sharp,\sqcup,\nabla\}$:}\hspace{0.5em}
  By repeated application of the inductive hypothesis (along with the preceding preservation lemmas for well-formedness and CFG/AI consistency), for each $n_i$ we may derive $\daigeval{\mathcal D_{i-1}}{M_{i-1}}{n_i}{v_i}{\mathcal D_i}{M_i}$, where $\mathcal D_0\triangleq\mathcal D$ and $M_0\triangleq M$.

  Then, either $\fnname\in \textsf{dom}(M_k)$, in which case we apply \textsc{Q-Match} to derive $$\daigeval{\mathcal D_0}{M_0}{n}{M_k(\fnname)}{\mathcal D_k[n\mapsto M_k(\fnname)]}{M_k}$$
  or $\fnname\not\in \textsf{dom}(M_k)$, in which case we apply \textsc{Q-Miss} to derive $$\daigeval{\mathcal D_0}{M_0}{n}{v}{D_k[v/n]}{M_k;\fnname\mapsto v}$$ where $v=f(v_1,\dots,v_k)$.

  \parindent=2em
  \hangindent=\parindent
  \underline{{\em Case} $f=\textsf{fix}$:}
  Since $\mathcal D\approxeq\langle L,E,\ell_0\rangle$, $c=n\gets \textsf{fix}(\underline\ell^{\;(k-1)},\underline\ell^{\;(k)})$.
  By the inductive hypothesis (along with the preceding preservation lemmas for well-formedness and CFG/AI consistency), $\daigeval{\mathcal D}{M}{\underline\ell^{\;(k-1)}}{v_1}{\mathcal D'}{M'}$ and $\daigeval{\mathcal D'}{M'}{\underline\ell^{\;(k)}}{v_2}{\mathcal D''}{M''}$.
  If $v_1=v_2$ then \textsc{Q-Loop-Converge} applies and we derive $\daigeval{\mathcal D}{M}{n}{v_1}{\mathcal D''}{M''}$.

  Otherwise, \textsc{Q-Loop-Unroll} applies but, since $n$ has more ancestors in $\textsf{unroll}(\mathcal D'',c)$ than in $\mathcal D$, we can not simply apply induction to derive the final premise.
  Note, however, that the final premise is also a query for $n$ in the unrolled DAIG, which by definition contains the edge $n\gets \textsf{fix}(\underline\ell^{\;(k)},\underline\ell^{\;(k+1)})$ and the empty reference cell $n[\varepsilon:\tau]$; therefore, it must be derived either by \textsc{Q-Loop-Unroll} or \textsc{Q-Loop-Converge}.
  If \textsc{Q-Loop-Unroll} is used then this argument repeats, yielding a tree of nested applications of \textsc{Q-Loop-Unroll} in direct correspondence to the sequence of abstract iterates $\underline\ell^{\;(k)},\underline\ell^{\;(k+1)},\dots$, which is an increasing sequence of abstract states computed by repeated application of $\nabla$.
  Therefore, by the convergence property of widening in the underlying abstract interpreter, this process may repeat only finitely many times before \textsc{Q-Loop-Converge} applies, yielding a finite derivation tree.

\end{proof}
\end{toappendix}




\section{Implementation and Evaluation}\label{sec:eval}

Here, we describe a prototype implementation and evaluation of demanded abstract interpretation via our DAIG framework.  Our evaluation studied two research questions:
\begin{itemize}[topsep=0.25em,leftmargin=1em]
  \item \textbf{Expressivity:} Does the DAIG framework allow for clean and straightforward implementations of rich analysis domains that cannot be handled by existing incremental and/or demand-driven frameworks?
  \item \textbf{Scalability:} For these rich analysis domains, what degree of performance improvement can be obtained by performing incremental and/or demand-driven analysis, as compared to batch analysis?
\end{itemize}



\subsection{Implementation}
Our DAIG framework is implemented in approximately 2,500 lines of OCaml code~\cite{dai-github-repo,dai-artifact-zenodo}. 
Incremental and demand-driven analysis logic, including demanded unrolling, operates over an explicit graph representation of DAIGs, but per-function memoization (i.e., the auxiliary memo table $M$ in the \cref{fig:query_semantics} semantics)
is provided via \code{adapton.ocaml}, an open-source implementation of the technique of \citet{DBLP:conf/oopsla/HammerDHLFHH15}.

\newsavebox{\SBoxModuleSig}
\begin{lrbox}{\SBoxModuleSig}\small
\begin{lstlisting}[language=caml,escapeinside={/*@}{@*/},morekeywords={sig,include,val,module}]
module type Dom = sig
  module Stmt : sig
    type t
    val skip : t
    include Adapton.Data.S with type t := t
  end
  type t                            /*@$\langle\Sigma^\sharp$,@*/
  val init : t                      /*@$\phantom{\langle}\varphi_0$,@*/
  val interpret : Stmt.t -> t -> t  /*@$\phantom{\langle}\denote\cdot^\sharp$,@*/
  val implies : t -> t -> bool      /*@$~~\sqsubseteq$,@*/
  val join : t -> t -> t            /*@$\phantom{\langle}\sqcup$,@*/
  val widen : t -> t -> t           /*@$\phantom{\langle}\nabla\rangle$@*/
  val is_bot : t -> bool
  include Adapton.Data.S with type t := t
end
\end{lstlisting}
\end{lrbox}

Our implementation is parametric in an abstract domain, and the effort required to instantiate the framework to a new abstract domain is comparable to the effort required to do so in a classical abstract interpreter framework.
The required module signature is essentially the abstract interpreter signature $\aitemplate$, extended with some standard utilities which can often be automatically derived.


\tightpara{Interprocedurality.}
Although the formalism is defined over control-flow graphs for clarity and brevity, our implementation supports context-sensitive analysis of non-recursive programs with static calling semantics (i.e., no virtual dispatch or higher-order functions).

In order to analyze such programs, we initially construct a DAIG only for the ``main'' procedure in the initial context.
Then, when a query is issued for the abstract state after a call, we construct a DAIG for its callee in the proper context.
When a query is issued at a location/context for which no DAIG has yet been constructed, we construct the DAIG for its containing function and analyze dataflow to its entry.

These operations are parametric in a context-sensitivity policy for choosing a context in which to analyze a callee at a call site.
Our implementation includes functors that implement context-insensitivity and also 1- and 2-call-site-sensitivity~\cite{sharir1978two}.

\subsection{Expressivity}\label{sec:eval:expressivity}
To demonstrate the expressivity of our DAIG framework, we have instantiated it with three existing well-known abstract interpretation techniques --- interval, octagon and shape analysis --- all of which are inexpressible using existing incremental and/or demand-driven analysis frameworks.
Here, we describe our experience applying the DAIG-based interval and shape analyses to a small set of programs. 
In \cref{sec:scalability}, we use the octagon domain to investigate the scalability of demanded analysis on synthetic benchmarks.

Together, these analysis implementations provide evidence for our approach's agnosticity to the underlying abstract domain, including domains with black-box external dependencies and/or complicated non-monotone abstract operations.

\tightpara{Interval Analysis.}
The interval abstract domain is a textbook example of an infinite-height lattice, requiring widening to guarantee analysis convergence.
An interval $[l,u]$ abstracts the set of numbers between lower bound $l$ and upper bound $u$.  Join and widen operations and abstract states are defined in the standard way~\cite{DBLP:conf/popl/CousotC77}.
Abstract interpretation in this domain is known as interval analysis and is commonly used, e.g., to verify the safety of array accesses. 
Interval analysis has been applied at industrial scale, for example by \citet{DBLP:conf/esop/CousotCFMMMR05}.

In practice, it is common to use an optimized off-the-shelf interval abstract domain such as that of APRON~\cite{DBLP:conf/cav/JeannetM09} or Elina~\cite{DBLP:conf/popl/SinghPV17}.
We have implemented an APRON-backed interval analysis for JavaScript programs in the DAIG framework.
As an indication of the flexibity of our framework, we were able to use the APRON library {\em without modification}.

In order to validate our implementation, we analyzed 23 array-manipulating programs --- with functions such as \texttt{contains}, \texttt{equals}, \texttt{swap}, and \texttt{indexof} --- from the test suite of Buckets.JS, a JavaScript data structure library~\cite{buckets}.

Using the 2-call-string-sensitive context policy, our analysis verified the safety of all 85 array accesses in the programs; with 1-call-string-sensitivity, it verified 71/74 (96\%), and with context-insensitive analysis it verified 4/18 (22\%).
These figures show that standard numerical analyses on DAIGs behave as they would in a batch analysis engine.

\tightpara{Shape Analysis.}
Precise analysis of recursive data structures such as linked lists is essential in many domains.
Such analysis relies on complex abstract domains that cannot be expressed in existing frameworks for incremental and demand-driven analysis.
We have implemented a DAIG-based demanded shape analysis for singly-linked lists.
An abstract state in this shape domain is a triple consisting of
\begin{itemize}[topsep=0.25em,leftmargin=1 em,itemindent=0.5em]
\item[-] A separation logic formula over points-to ($\alpha.f\mapsto \alpha'$) and list-segment (\texttt{lseg}$(\alpha,\alpha')$) atomic propositions, stating respectively that the $f$ field of the object at symbolic address $\alpha$ points to $\alpha'$ and that there exists a sequence of \texttt{next} pointer dereferences from $\alpha$ to $\alpha'$~\cite{DBLP:conf/lics/Reynolds02},
\item[-] A collection of pure constraints: equalities and disequalities over memory addresses, and
\item[-] An environment mapping variables to memory addresses.
\end{itemize}
Join, widen, and implication all rely on a collection of rewrite rules over such states from \citet{DBLP:conf/sas/ChangRN07} (specialized to a fixed inductive definition for list segments).
All told, the implementation of this shape domain requires approximately 500 lines of OCaml code.

We have applied this DAIG-based shape analysis to successfully verify the correctness and memory-safety of the list \code{append} procedure of \cref{fig:overview:cfg}, along with several linked list utilities from the aforementioned Buckets.js library including \code{foreach} and \code{indexof}~\cite{buckets}.
Analysis of the $\ell_3$-to-$\ell_4$-to-$\ell_3$ loop of the list \code{append} procedure converges in one demanded unrolling with a precise result.

\subsection{Scalability}\label{sec:scalability}
To study scalability, we conducted an initial investigation of what performance improvements are possible with demanded analysis variants in our framework.
We compared the performance of analysis with and without incrementality and demand on interleaved sequences of program edits and queries.
Throughout this section, our framework is instantiated with a context-insensitive APRON-backed octagon domain: a relational numerical domain representing invariants of the form $\pm x \pm y \leq y$, widely used in practice due to its balance of expressivity and efficiency~\cite{DBLP:journals/lisp/Mine06}.

\begin{figure*}
\includegraphics[width=0.33\textwidth]{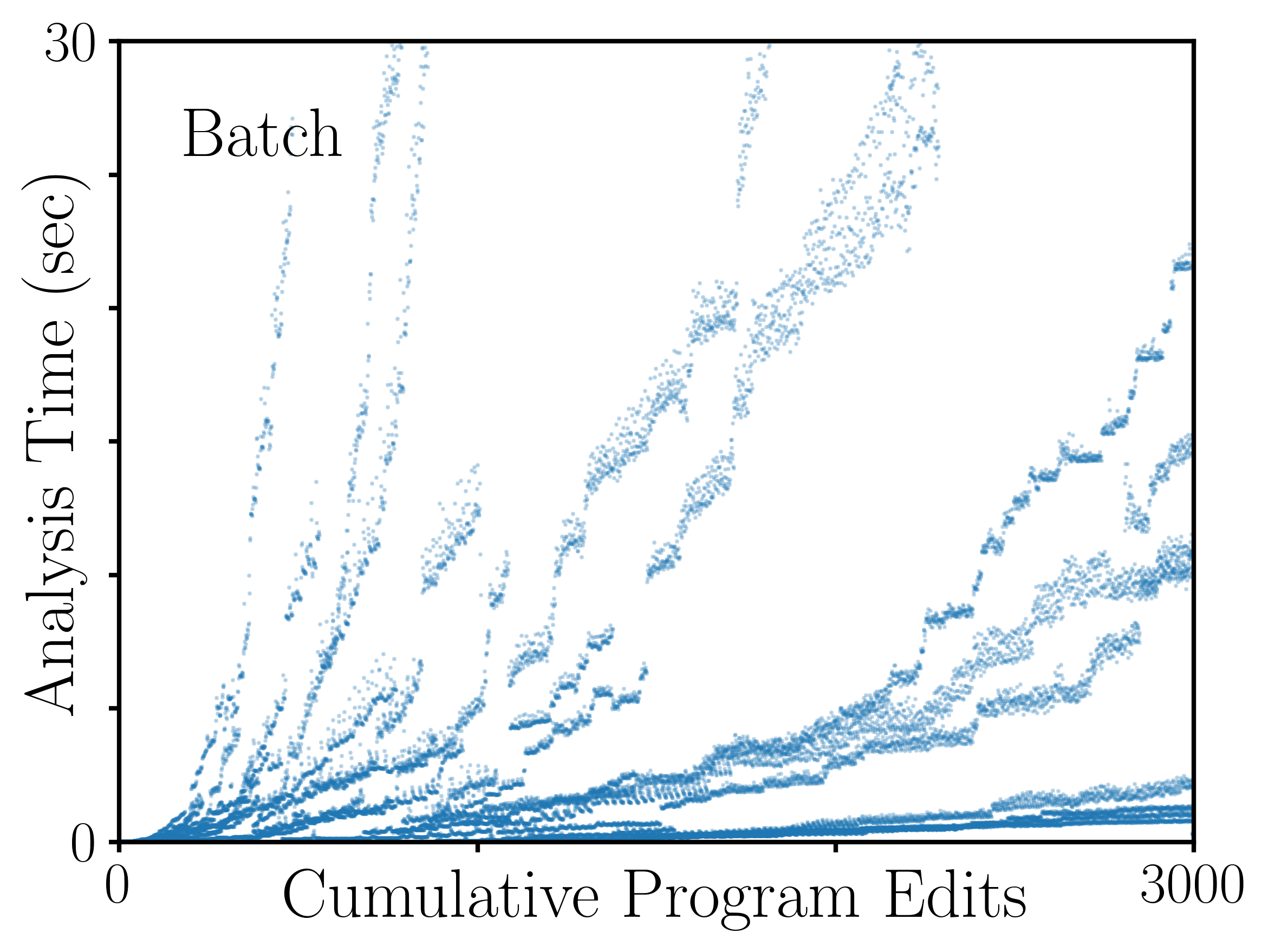}
\includegraphics[width=0.33\textwidth]{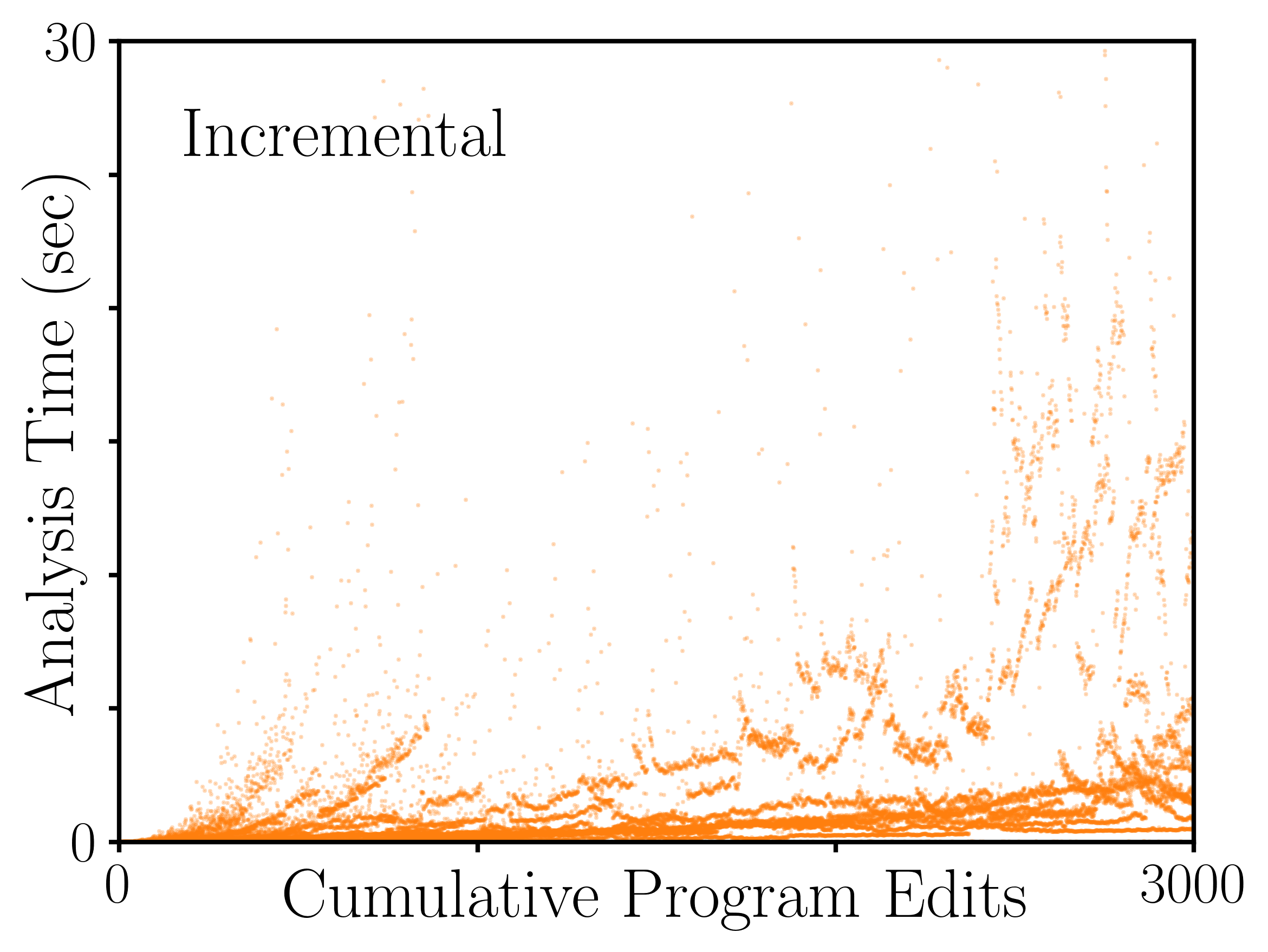}
\includegraphics[width=0.318\textwidth]{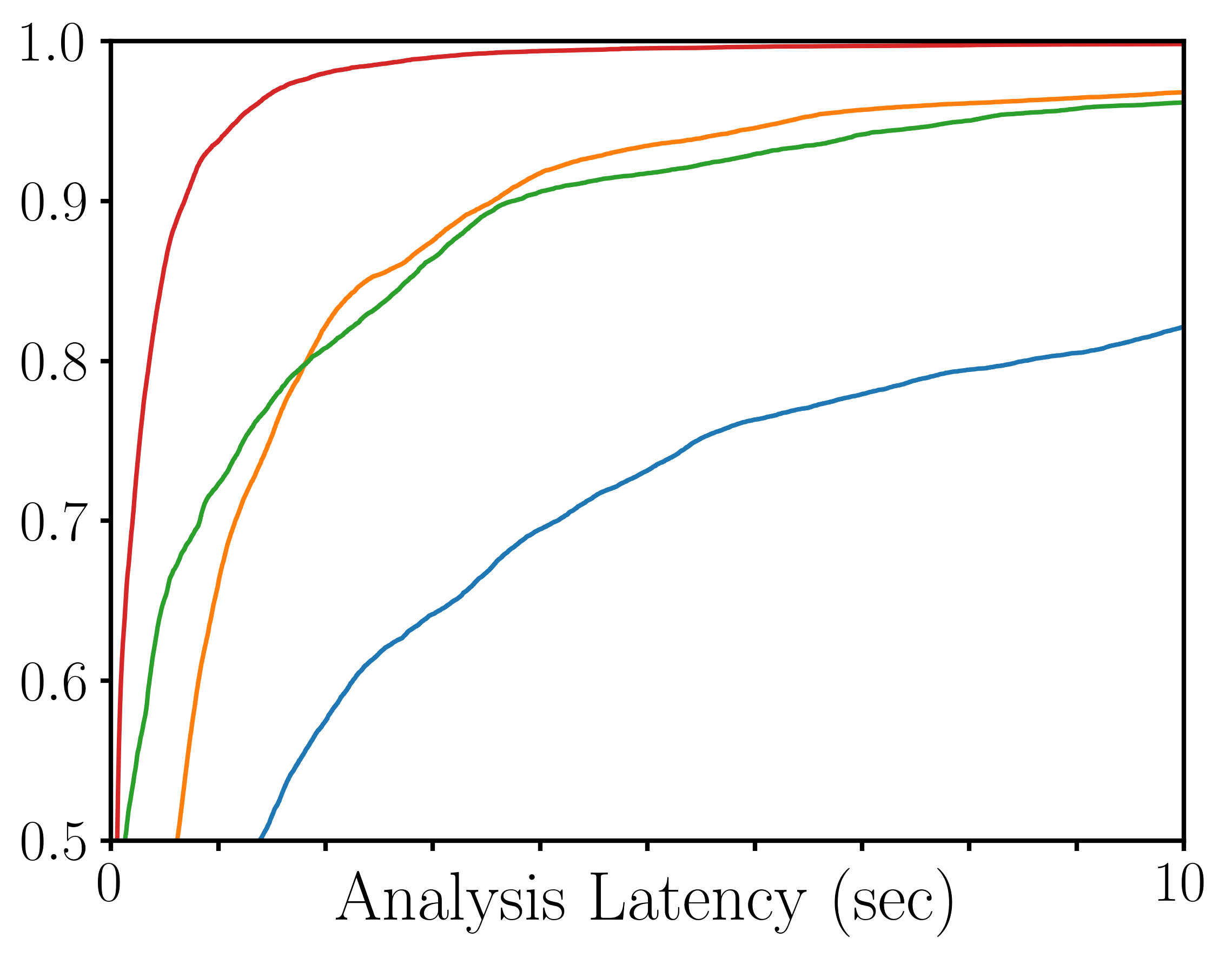} 
\hfill
\includegraphics[width=0.33\textwidth]{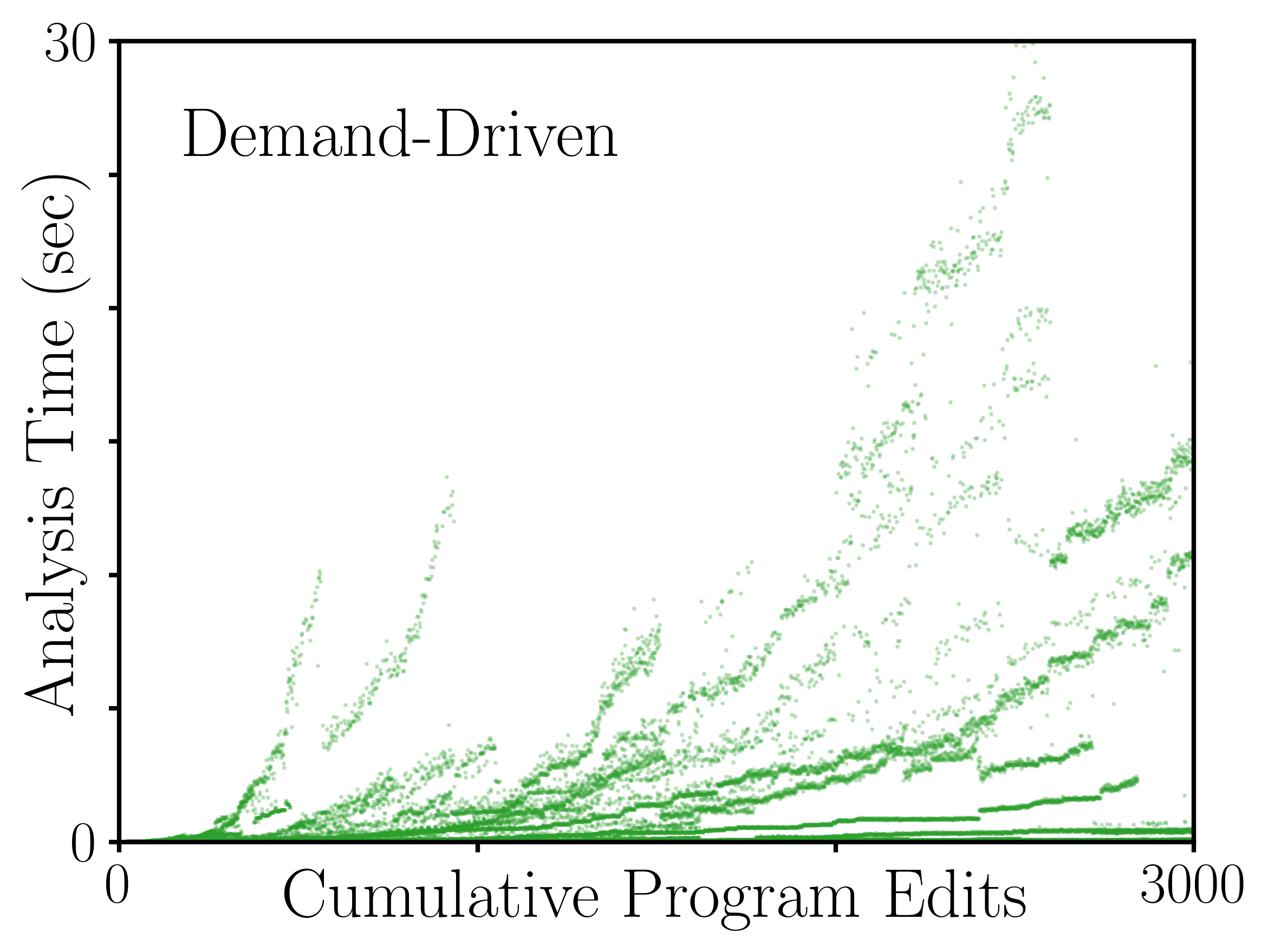}
\includegraphics[width=0.33\textwidth]{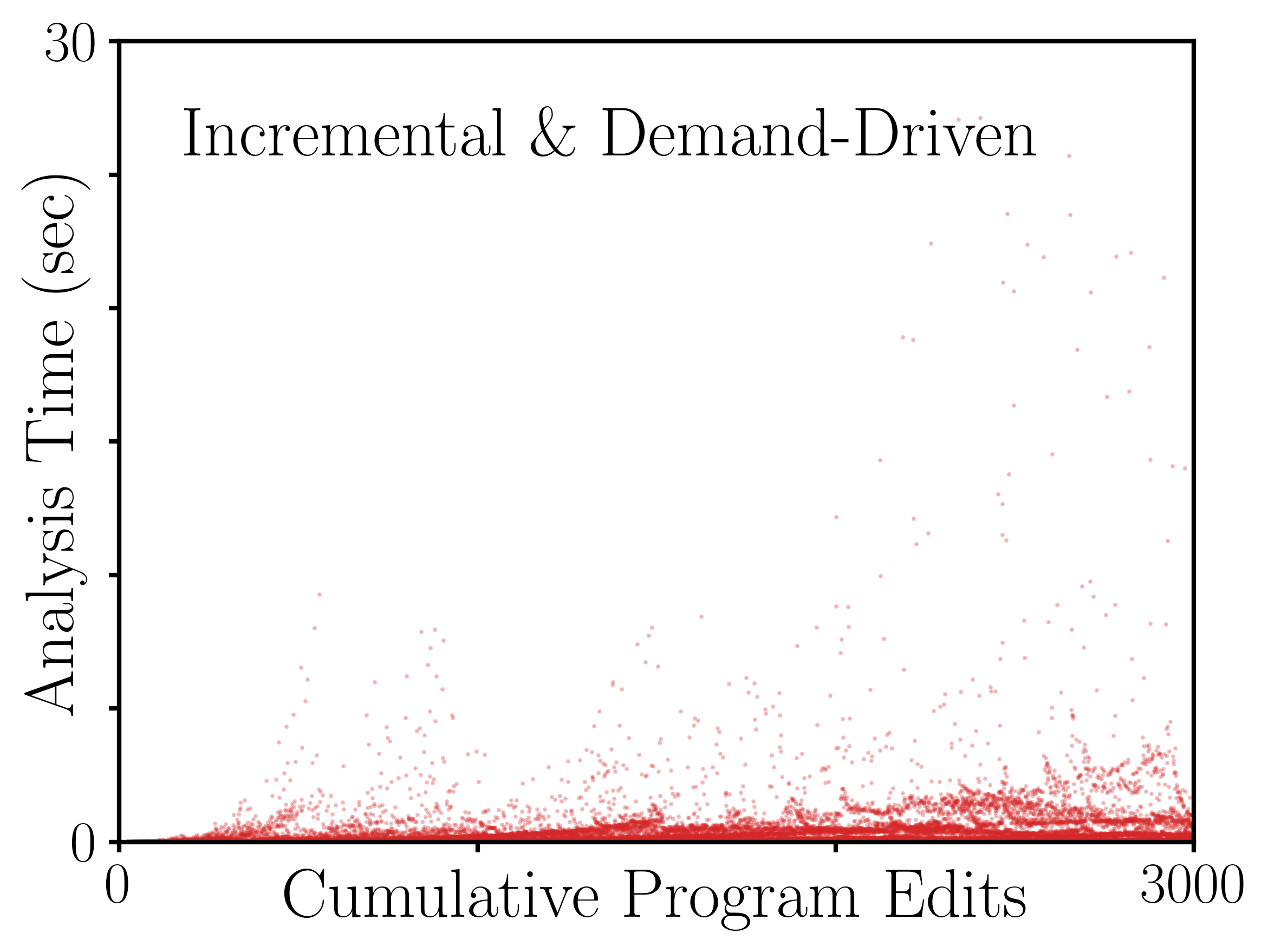}
\hfill
\raisebox{9ex}{\small
  \fbox{\begin{tabular*}{0.30\textwidth}[b]{@{\extracolsep{\fill}}lrrrrr@{}}
    & \multicolumn{5}{c}{Analysis Time (sec)} \\
    & mean & p50 & p90 & p95 & p99  \\ \midrule
    Batch & 9.0 & 1.4 & 18.9 & 36.2 & 173.6 \\
    Incr. & 1.7 & 0.6 & 3.6 & 6.3 & 16.6 \\
    DD & 1.5 & 0.1 & 3.7 & 7.9 & 16.7 \\
    I\&DD & 0.3 & 0.1 & 0.7 & \demandedpninetyfive & 3.0
  \end{tabular*}}
}
%
  
  \caption[Evaluation]{Performance of octagon analysis on the synthetic workload of interleaved program edits and analysis queries described in \cref{sec:scalability}.
    The four scatter plots show the scaling of each configuration as the program size is increased by edits, and their color-coding serves as a legend to the fifth figure:
    a cumulative distribution plot showing the fraction of analysis runs ($y$ axis) completed by each configuration within some time interval ($x$ axis).
    Lastly, the table shows summary statistics for each configuration, including the mean, median, $90$th, $95$th, and $99$th percentile analysis latency.
  }
  \label{fig:eval:scalability}
\end{figure*}

To exercise the analyses, we created synthetic workloads consisting of 3,000 random edits to an initially-empty program.
Programs are generated in a JavaScript subset
with assignment, arrays, conditional branching, while loops, and (non-recursive) function calls of the form \code{x = f(y)}.
An ``edit'' is an insertion of a randomly generated statement, if-then-else conditional, or while loop at a randomly-sampled program location, with 85\%, 10\%, and 5\% probability respectively, and statements and expressions are generated probabilistically from their respective grammars.

We evaluate four analysis configurations on this workload:
\begin{itemize}[topsep=0.25em,leftmargin=0.5em,itemindent=1em]
\item[(1)] {\em Batch} analysis: Classical whole-program abstract interpretation, fully re-analyzing the entire program from scratch in response to each edit.
\item[(2)] {\em Incremental} analysis: An incremental-only configuration which applies the edit semantics to dirty as few previously-computed analysis results as possible, but eagerly recomputes all dirtied cells.
\item[(3)] {\em Demand-driven} analysis: A demand-driven-only configuration which dirties the full DAIG after each edit, but applies the query semantics to avoid computing analysis results that aren't demanded.
\item[(4)] {\em Incremental \& demand-driven} analysis: The full demanded abstract interpretation technique, which applies both the edit and query semantics to maximize reuse and minimize redundant computation.
\end{itemize}
In the demand-driven configurations, queries are issued at five randomly-sampled program locations between each edit.  Note that since the first three configurations were implemented atop our DAIG framework, which is designed to support both incremental and demand-driven analysis, they may not be as tuned as specialized implementations.

Each plot includes data points from 9 separate trials, with fixed random seeds such that the same edits (and, in the two demand-driven configurations, queries) are issued to each configuration.
In total, this data set includes 27,000 analysis executions in each exhaustive configuration and 135,000 queries in each demand-driven configuration.

The results, as shown in \cref{fig:eval:scalability}, indicate that while incremental and demand-driven analysis each significantly improve analysis latencies with respect to the batch analysis baseline, combining the two provides an additional large reduction in latency.
This effect is most apparent in the tail of the distribution, since edits that dirty large regions of the program are costly for incremental analysis, and queries that depend on large regions of the graph are costly for demand-driven analysis.
By combining incremental dirtying with demand-driven evaluation, demanded abstract interpretation mitigates these worst-case scenarios and consistently keeps analysis costs low even as the program grows.

In particular, at the 95$^\mathit{th}$ percentile, the \demandedpninetyfive s latency of incremental demand-driven analysis is more than five times lower than the next best configuration, and potentially low enough to support interactive use.  \cref{fig:eval:scalability} gives a cumulative distribution of analysis latencies, again showing the large advantage of the incremental demand-driven analysis over other configurations.

\newpage
\section{Related Work}

\tightpara{Incremental Computation.}
Techniques for the  efficient caching and reuse of computation results, particularly those based on
memoization of pure functions~\cite{DBLP:conf/icfp/AbadiLL96,DBLP:conf/lfp/FieldT90,DBLP:conf/popl/PughT89}
and dependency graphs~\cite{DBLP:conf/popl/DemersRT81,DBLP:conf/popl/Reps82},
have been the subject of a great deal of research and seen widespread practical application.

More recently, dependency graph-based approaches to incremental computation have improved on generic memoization and graph-based techniques, allowing for fine-grained automatic caching and reuse even in the presence of changes to inputs or an underlying data store~\cite{DBLP:conf/popl/AcarBH02,DBLP:conf/popl/AcarAB08}.
Building on these graph-based techniques for self-adjusting computation, some recent work has focused on support for interactive and demand-driven computations~\cite{DBLP:conf/oopsla/HammerDHLFHH15,DBLP:conf/pldi/HammerKHF14}.
Although this approach yields a general and powerful system for incremental computation, its low-level primitives make it difficult to express the complex fixed-point computation over cyclic control-flow graphs in arbitrary abstract interpretations.
We take inspiration from demanded computation graphs but instead specialize the language of demanded computations to demanded abstract interpretations, both with syntactic structures and with a query/edit semantics which dynamically modifies the dependency graph to model such computations.

\tightpara{Incremental Analysis.}
The application of incremental computation to program analysis is similarly well-studied, going back at least to the development of incremental data\-flow analyses to support responsive continuous compilation~\cite{DBLP:conf/popl/Ryder83,DBLP:conf/sigplan/Zadeck84}.
Recent work has contributed incremental versions of several classes of program analysis, including IFDS/IDE dataflow analyses~\cite{DBLP:conf/icse/ArztB14, DBLP:conf/issta/DoALBSM17} and analyses based on extensions to Datalog~\cite{DBLP:conf/kbse/SzaboEV16,DBLP:journals/pacmpl/SzaboBEV18}.
These specialized approaches offer effective solutions for certain classes of program analysis, but place restrictions on abstract domains that rule out arbitrary abstract interpretations in infinite-height domains. 

Compositional program analysis, in which summaries are computed for individual files or compilation units rather than a whole program, naturally supports incrementality in the sense that results need only be recomputed for changed files.  This has shown to be very effective for scaling program analyses to massive codebases in CI/CD systems~\cite{DBLP:conf/nfm/CalcagnoD11,DBLP:journals/cacm/DistefanoFLO19,DBLP:conf/foveoos/FahndrichL10}, but it operates at a much coarser granularity than both the aforementioned approaches and our own, since it is designed to scale up to massive programs rather than to minimize analysis latencies at development-time.

\citet{DBLP:conf/cav/LeinoW15} propose a fine-grained incremental verification technique for the Boogie language, which verifies user-provided specifications of imperative procedures. These specifications include loop invariants, allowing their algorithm to ignore cyclic dependencies altogether.

\tightpara{Demand-Driven Analysis.}
Demand-driven techniques for dataflow analysis are also well-studied.  The intra-pro\-ce\-dur\-al problem was studied by \citet{10.1007/BF00264319}.  Several extensions to inter-procedural analysis have been presented, for example, by \citet{Reps1994SolvingDV}, \citet{DBLP:conf/popl/DuesterwaldGS95}, and \citet{DBLP:journals/tcs/SagivRH96}.  In nearly all cases previous work has been focused on finite domains.  The work of \citet{DBLP:journals/tcs/SagivRH96} allows for infinite domains of finite height, but does not consider infinite-height domains like intervals.

Any static analysis expressible as a context-free-language reachability (CFL-reachability) problem can be computed in a demand-driven fashion as a ``single-source'' problem~\cite{reps98cfl}.  As such, a number of papers have presented demand-driven algorithms for flow-insensitive pointer analysis~\cite{DBLP:conf/pldi/HeintzeT01,sridharan05demand,DBLP:conf/ecoop/SpathDAB16}.

Reference attribute grammars (RAGs) are declarative specifications of properties over ASTs (including potentially-cyclic flow analyses) which can be evaluated incrementally and on-demand~\cite{DBLP:journals/scp/MagnussonH07,incremental_rag_tr}.
Termination of RAG evaluation requires that all cyclic computations converge to a fixed-point in finitely-many iterations~\cite{DBLP:journals/scp/MagnussonH07,DBLP:conf/sigplan/Farrow86}; this convergence property holds for finite domains with monotone operators but may also be achieved through other means (e.g. widening).

Improving on prior work, our framework comes with proofs of termination and from-scratch consistency, and specifies the exact conditions required to ensure termination in infinite-height domains with non-monotone widening operators.

\section{Conclusion}

We have presented a novel framework for demanded abstract interpretation, in which an arbitrary abstract interpretation can be made both incremental and demand-driven.  Unlike previous frameworks, ours supports arbitrary lattices and widening operators.  The framework is based on a novel demanded abstract interpretation graph  (DAIG) representation of the analysis problem, where careful handling of loops ensures the DAIG remains acyclic.  We have proved various key properties of the framework, including soundness, termination, and from-scratch consistency.  Our implementation shows that complex analyses can be easily implemented with our framework, with the potential for significant performance wins in incremental and demand-driven scenarios.

\begin{acks}                            
  We thank Matthew A. Hammer and Jared Wright for their valuable contributions in the early stages of this research.
  We also thank the anonymous reviewers and members of the CUPLV lab for their helpful reviews and suggestions.
  This research was supported in part by the National Science Foundation under grants CCF-1619282, CCF-2008369, and CCF-2007024, and also by a gift from Oracle Labs.

\end{acks}

\bibliography{conference.short,refs.short}

\end{document}